\documentclass[useAMS,usenatbib]{mn2e}
\usepackage{graphicx,amsmath,multirow,amssymb}
\usepackage{subfig}
\usepackage{natbib}
\newcommand{\comment}[1]{}

\def\simgt{\lower.5ex\hbox{$\; \buildrel > \over \F sim \;$}}
\def\simlt{\lower.5ex\hbox{$\; \buildrel < \over \sim \;$}}

\title[]{AGB dust and gas ejecta in extremely metal-poor environments}
\author[Dell'Agli et al.]{F. Dell'Agli$^{1,2}$, R. Valiante$^3$, D. Kamath$^{4,5}$, P. Ventura$^3$,
D. A. Garc\'{\i}a--Hern\'andez$^{1,2}$\\
$^{1}$Instituto de Astrof\'{\i}sica de Canarias (IAC), E-38200 La Laguna, Tenerife, Spain \\
$^{2}$Departamento de Astrof\'{\i}sica, Universidad de La Laguna (ULL), E-38206 La Laguna, Tenerife, Spain \\
$^3$INAF -- Osservatorio Astronomico di Roma, Via Frascati 33, 00040, Monte Porzio Catone (RM), Italy \\
$^4$Department of Physics \& Astronomy, Macquarie University, Sydney, NSW 2109, Australia  \\
$^5$Astronomy, Astrophysics and Astrophotonics Research Centre, Macquarie University, Sydney, NSW 2109, Australia \\
}

\begin{document}

\date{Accepted, Received; in original form }

\pagerange{\pageref{firstpage}--\pageref{lastpage}} \pubyear{2012}

\maketitle

\label{firstpage}

\begin{abstract}
We present asymptotic giant branch (AGB) models of metallicity $Z=10^{-4}$ and $Z=3\times 10^{-4}$,
with the aim of understanding how the gas enrichment and the dust production change in 
very metal-poor environments and to assess the general contribution of AGB stars to the
cosmic dust yield. The stellar yields and the dust produced are determined by the change in
the surface chemical composition, with a transition occurring at $\sim 2.5~M_{\odot}$. Stars
of mass $M < 2.5~M_{\odot}$ reach the carbon stage and produce carbon dust, whereas their 
higher mass counterparts produce mainly silicates and alumina dust; in both cases the
amount of dust manufactured decreases towards lower metallicities. The $Z=10^{-4}$ models show a
complex and interesting behaviour on this side, because the efficient destruction of the
surface oxygen favours the achievement of the C-star stage, independently of the initial mass.
The present results might indicate that the contribution from this class of stars to the
overall dust enrichment in metal-poor environments is negligible at redshifts $z>5$.
\end{abstract}


\section{Introduction}
The stars with initial mass in the range $1-8~M_{\odot}$ after the core He-burning phase
evolve through the asymptotic giant branch (the thermally-pulsing phase, hereafter mentioned as AGB, Iben 1981). Despite this evolutionary
phase being only a few percent of the whole stellar life, it is of great importance,
because it is during the AGB evolution that the 
stars lose the majority of their mass, thus contributing to the pollution of the
interstellar medium.

Among others, understanding the gas pollution from AGB stars is crucial to investigate the
chemical evolution of the Milky Way (e.g. Romano et al. 2010, Kobayashi et al. 2011, 
Ginolfi et al. 2018), Local Group galaxies (e.g. Schneider et al. 2014, Vincenzo et al. 2016), 
the interstellar medium of galaxies (e.g. Romano et al. 2017), up to very high redshift 
(e.g. Mancini et al. 2015), the formation of multiple 
populations in globular clusters (e.g. D'Ercole et al. 2008). For example, the importance of 
studying the yields of the CNO elements from AGB stars has been recently outlined 
by \citet{vincenzo18a}, in a paper focused on the reconstruction of the star formation history (SFH) of 
galaxies, based on the CNO trends observed in the interstellar medium. 
On the other hand, \citet{vincenzo18b} showed the importance of taking into account all 
the stellar sites of CNO nucleosynthesis, including AGB stars, to perform detailed 
chemo-dynamical simulations on cosmological scales, in which the N$/$O vs O$/$H trends
are used to reconstruct the evolution of galaxies with redshift.

A further reason for the interest towards AGB stars is that they have been proposed as 
important dust manufacturers, owing to the
thermodynamic conditions in their wind, which are extremely favourable to allow
condensation of gaseous molecules into solid particles \citep{gail99}.
However, the role played by AGB stars as cosmic dust producers has still to be fully
understood: contrary to early investigations, more recent studies
outlined an important contributions from AGBs event at high redshifts 
\citep{valiante09, valiante11, valiante17}. The determination of the amount
of dust produced by AGB stars and the corresponding size distribution function is
required to calculate the extinction properties associated with dust grains, 
which is a fundamental information to interpret the optical-near-infrared properties 
of high-z quasars and gamma-ray burst spectra (Maiolino et al. 2004; Gallerani et al. 2010).

In the recent past, the modelling of the AGB phase has made significant steps forwards,
with the inclusion of the description of the dust formation mechanism within the stellar
evolution framework \citep{ventura12a, ventura12b, nanni13, nanni14}. These models
have been extensively tested against near- to mid-infrared observations, primarily obtained
with \emph{Spitzer}, of AGB stars in the Magellanic Clouds \citep{flavia14, flavia15a, 
flavia15b, nanni16, nanni18} and in Local Group galaxies 
\citep{flavia16, flavia18a, flavia19}. The results from the \emph{Gaia} data release 2 
allowed the comparison with AGB stars in the LMC \citep{ambra}.
These studies will receive a strong boost after
the launch of the \emph{James Webb} Space Telescope, which will provide extensive 
infrared (IR) data
for the AGB population of all the galaxies in the Local Group, and possibly beyond.

The above mentioned studies required use of AGB models, including the description of
dust formation, with metallicities ranging to $Z=10^{-3}$ to the metallicity typical
of LMC stars, i.e. $Z=8\times 10^{-3}$. Solar and supersolar models are also available in the literature \citep{nanni13,nanni14,flavia17, ventura18,zhukovska13}.

In this work, we focus our attention towards the low-metallicity tail and present
AGB models, of initial mass in the range $1-7~M_{\odot}$, down to the metallicity $Z=10^{-4}$. We calculate the gas yields produced by these stars during the AGB life, making them available  
for the studies focused on galaxy evolution, which require the knowledge of the 
contribution from low-metallicity AGB stars to the chemical enrichment, particularly
for what attains carbon and nitrogen.

Furthermore, we compute the dust manufactured by these stars during the AGB life.
In a cosmological context, our goal is to address the fundamental issue of 
the contribution of AGB stars to the dust enrichment in the early Universe, something
that requires the quantification of the dust produced by AGB stars in very 
metal-poor environments.
Our aim is to find some trends with metallicity, down to the point where a straight
extrapolation of the results is sufficient to infer how dust formation works in
metal-poor environments.

The paper is structured as follows: in section 2 we describe the codes used to model
stellar evolution and dust formation; a detailed discussion of the main physical and 
chemical properties of the stars of metallicity $Z=10^{-4}$ and $Z=3\times 10^{-4}$
are given in section 3; sections 4 and 5 provide a general description of the 
behaviour of, respectively, the variation of the surface chemical composition and 
of dust production, in AGB stars of metallicity $Z \leq 4\times 10^{-3}$; in section 6 
we discuss the contribution of metal-poor AGB stars to the cosmic dust yield; the 
conclusions are given in section 7.

\section{Stellar evolution and dust formation modelling}
\label{ATON}
The evolutionary sequences used in the present analysis were started from the pre-main 
sequence and followed until the final AGB stages, when the convective envelope was almost 
entirely consumed. 

\subsection{Initial chemical composition}
\label{inchim}
For this study we calculated new evolutionary sequences, of metallicity $Z=3\times 10^{-4}$,
and mass $M\leq 2.5~M_{\odot}$. These models complete the higher mass models of the same
metallicity by \citet{marcella13}. We also calculated ex novo a full set of models of
metallicity $Z=10^{-4}$. 

To discuss the trend of the chemical and dust properties with the metallicity,
we considered in the present work models already published by our group. For the 
metallicity $Z=10^{-3}$ we used the models published in \citet{ventura14b}; for 
$Z=2\times 10^{-3}$ we relied on the results by \citet{ventura16a} (the dust was 
calculated for the present scope). In the newly calculated
$Z=10^{-4}$ and $Z=3\times 10^{-4}$ sequences and in the $Z=10^{-3}$ and $Z=2\times 10^{-3}$ 
models we used an $\alpha-$enhanced mixture with $[\alpha/$Fe$]=+0.4$, taken from \citet{gs98}. 
We also discuss $Z=4\times 10^{-3}$ models, published in \citet{ventura14a}; because 
this study is focused on stars in the cluster 47 Tuc, the $\alpha-$enhancement considered 
is  $[\alpha/$Fe$]=+0.2$.

\subsection{Stellar evolution modelling}
Stellar evolution models were calculated by means of the code {\sc ATON}.
An exhaustive discussion of the numerical structure of the code, with all the 
micro-physics input used, can be found in \citet{ventura98}; the latest updates are given 
in \citet{ventura09}. For what concerns the choices regarding the macro-physics adopted, 
we discuss here the input most relevant for the present work:

\begin{enumerate}

\item{{\it Convection.} The convective instability was treated by means of the Full
Spectrum of Turbulence (hereafter FST) model for turbulent convection \citep{cm91}. 
This choice is particularly relevant for the mass domain $M > 3~M_{\odot}$, because
use of the FST description leads to a thermodynamical stratification of the envelope
extremely favourable to the ignition of Hot Bottom Burning (HBB), which consists in the
activation of a proton-capture nucleosynthesis at the base of the convective
envelope \citep{vd05}.

In nuclearly active regions unstable to convection, we used a diffusive-like description,
where nuclear burning and mixing of chemical are self-consistently coupled. The set of
equations used were taken by \citet{eoll}. Convective
velocities are allowed to decay exponentially from the border of convective core
during the H- and He-burning phases, with an e-folding distance of $0.02~H_p$;
during the TP-AGB phase we consider that the velocity of the convective eddies 
in the shell which develops at the ignition of each thermal pulse (hereafter TP) or 
within the convective envelope decay within radiatively stable regions with 
an e-folding distance of $0.002~H_p$.}

\item{{\it Mass loss.} 
For oxygen-rich AGB stars, the mass loss was modelled according to \citet{blocker95}. The 
free parameter entering this recipe was set to $\eta_R = 0.02$. For carbon stars we
use the description of mass loss by the Berlin group \citep{wachter02, wachter08}.
}

\item{{\it Radiative opacities.} The radiative opacities for temperatures above $10^4$ K 
were calculated using the OPAL online tool (Iglesias \& Rogers 1996); for smaller 
temperatures we used the {\sc AESOPUS} tool described in Marigo \& Aringer (2009). 
This choice allows to account for the increase in the opacity associated with the 
change in the surface chemistry determined by inwards penetration of the convective envelope.}

\end{enumerate}

\begin{figure*}
\begin{minipage}{0.48\textwidth}
\resizebox{1.\hsize}{!}{\includegraphics{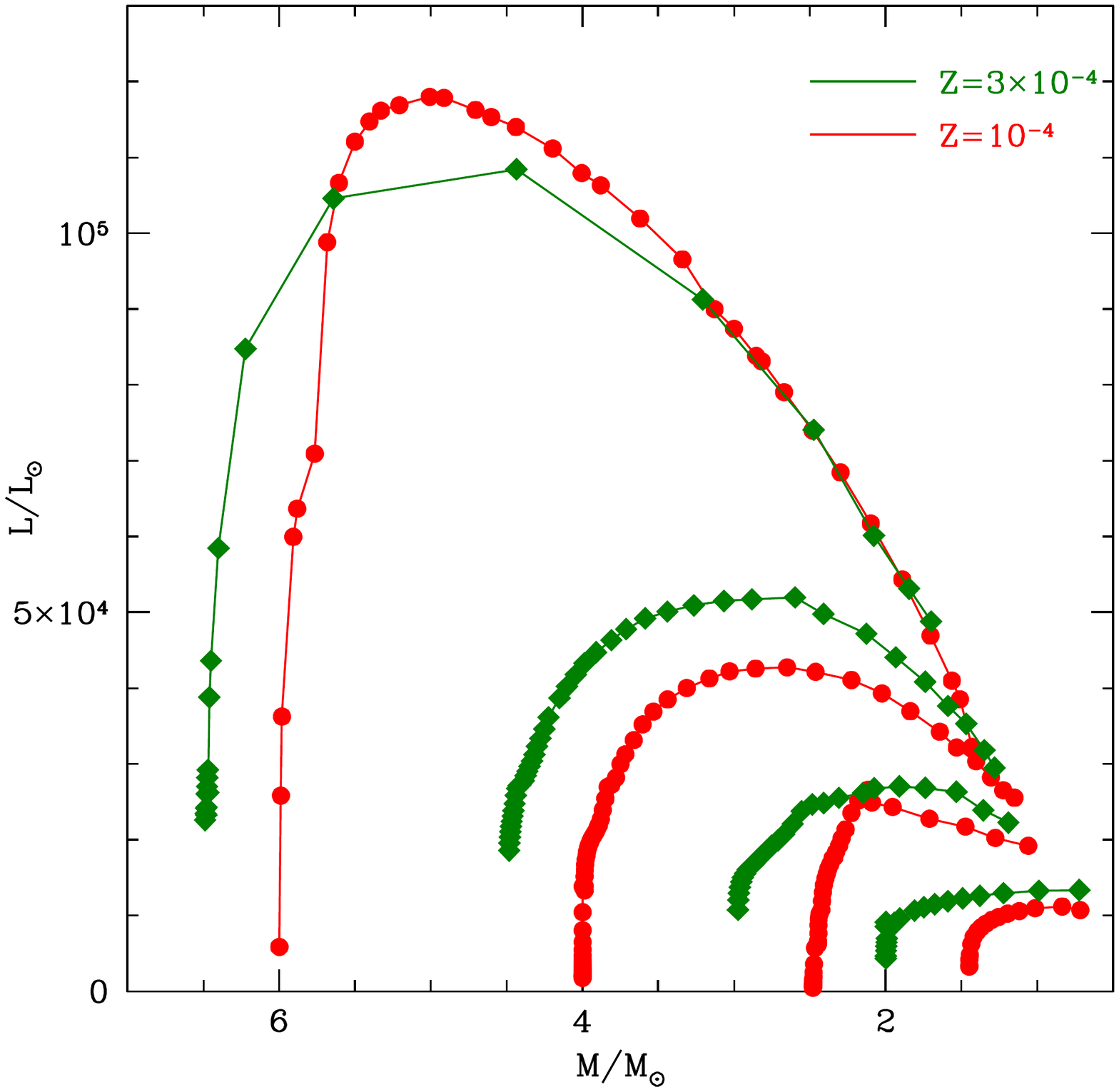}}
\end{minipage}
\begin{minipage}{0.48\textwidth}
\resizebox{1.\hsize}{!}{\includegraphics{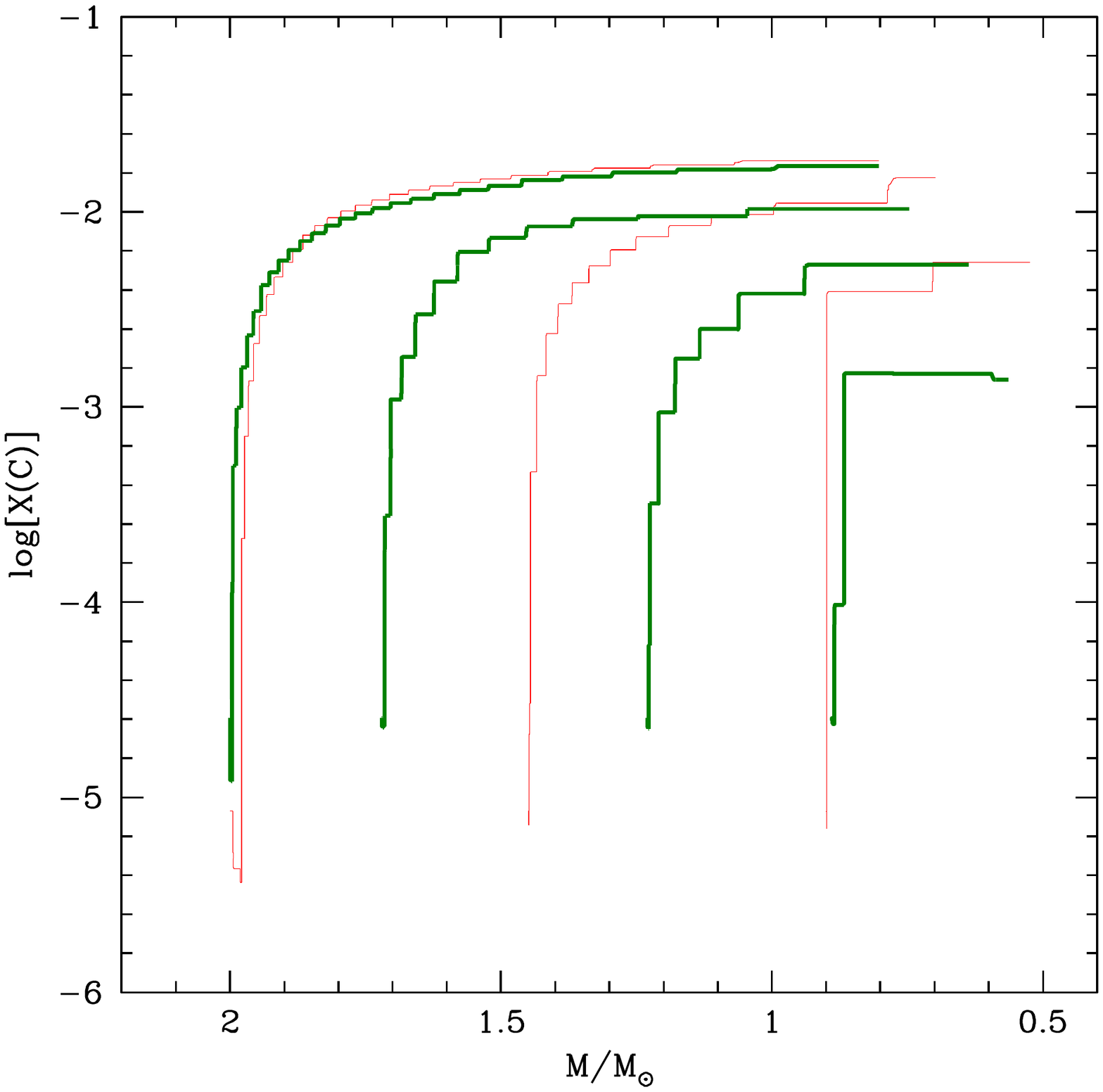}}
\end{minipage}
\vskip-80pt
\begin{minipage}{0.48\textwidth}
\resizebox{1.\hsize}{!}{\includegraphics{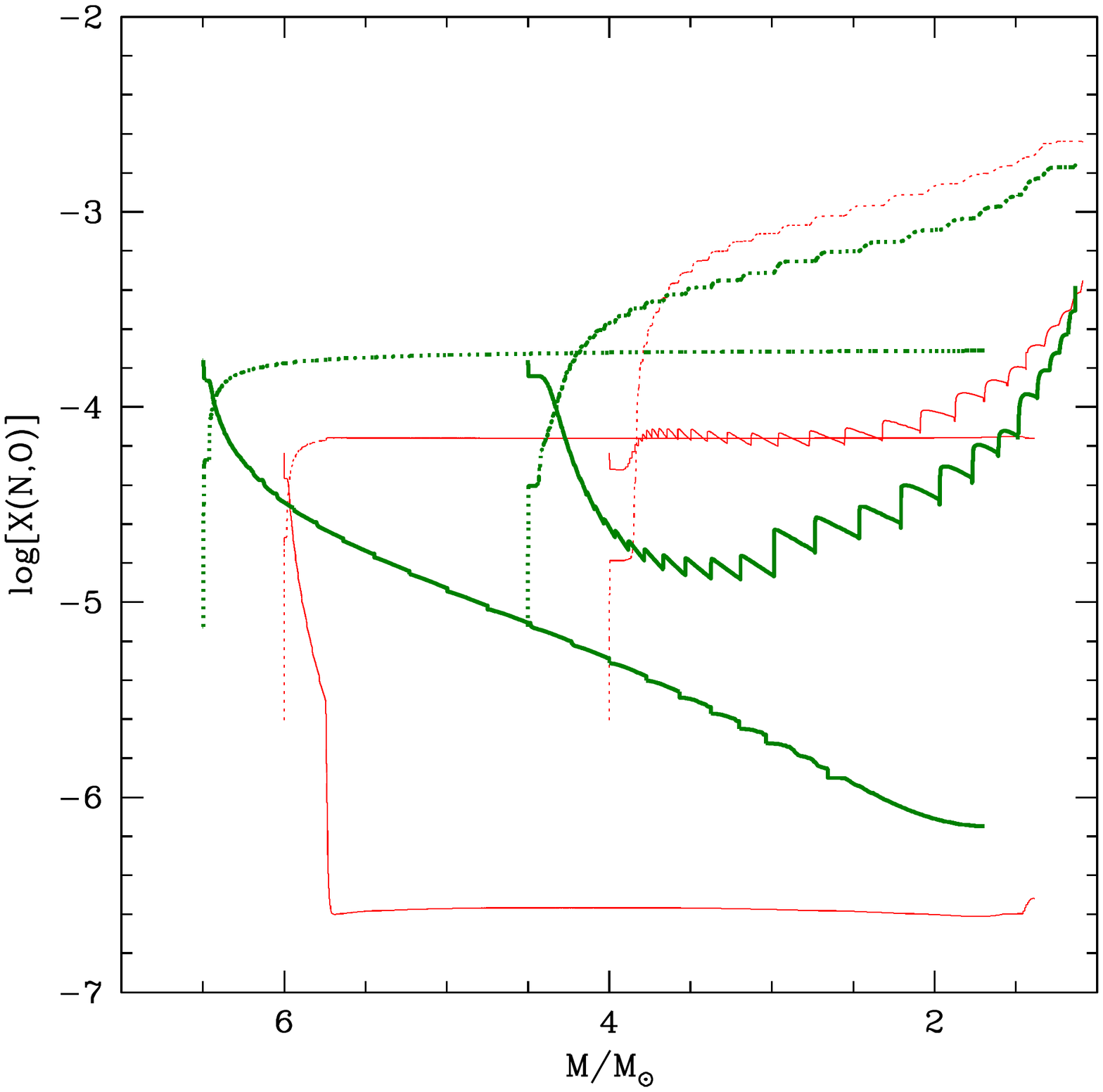}}
\end{minipage}
\begin{minipage}{0.48\textwidth}
\resizebox{1.\hsize}{!}{\includegraphics{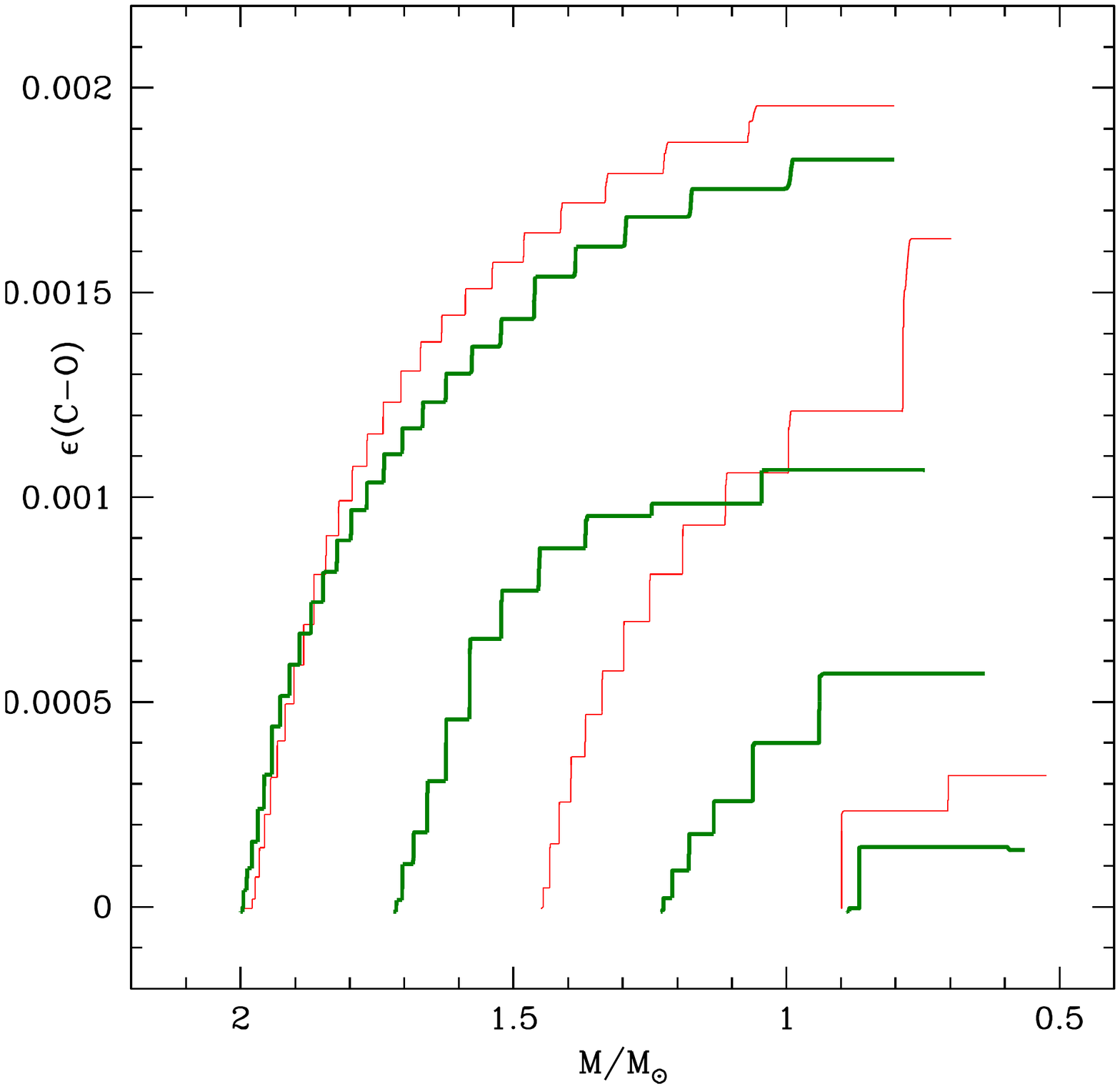}}
\end{minipage}
\vskip-60pt
\caption{Top, left: the AGB evolution of the luminosity of AGB stars of different initial mass and
metallicity $Z=10^{-4}$ (red points) and $Z=3 \times 10^{-4}$ (green diamonds), as a function 
of the current mass of the star. The different points indicate the values of the luminosity 
in the middle of each inter-pulse phase. The panels in the right show the evolution of the 
surface carbon mass fraction (top) and of the excess of carbon with respect to oxygen (bottom)
of models of different mass and metallicity $Z=10^{-4}$ (thin red solid line) and $Z=3 \times 10^{-4}$ (thick green solid line); the initial mass can be deduced by the abscissa of the starting point of
each track. Bottom, left: the evolution of the surface abundance of oxygen (solid lines) and
nitrogen (dotted lines) of $Z=10^{-4}$ models of initial mass $4~M_{\odot}$ and $6~M_{\odot}$
(thin red tracks) and of $Z=3\times 10^{-4}$ stars of initial mass $4.5~M_{\odot}$ and 
$6.5~M_{\odot}$ (thick green tracks).}
\label{fmodels}
\end{figure*}

\subsection{The description of dust formation}
\label{dustinput}
To describe dust production we apply the description of the wind
of AGB stars proposed by the Heidelberg group \citep{fg01, fg02, fg06}, which allows to 
determine the size of the solid particles formed in the circumstellar envelope, for the various
dust species considered. This method was applied in the past works on this argument
by our group \citep{ventura12a, ventura12b, marcella13, ventura14b, flavia17, ventura18} and
by the Padua team \citep{nanni13, nanni14, nanni16, nanni18}.

The choice of the dust species considered is based on stability arguments of the
different compounds \citep{sharp90}. Because the CO molecule is extremely stable, it is assumed
that in C-rich environments no oxygen is left to form dust, and that the species formed
are silicon carbide (SiC) and amorphous carbon. Conversely, in oxygen-rich envelopes,
no carbon-bearing dust grains can form, thus leaving space for the formation of
alumina dust (Al$_2$O$_3$) and silicates only. The latter species is split into the three
components olivine (Mg$_2$SiO$_4$), pyroxene (MgSiO$_3$) and quartz (SiO$_2$); olivine
is the most stable, thus we will often refer to it in the following for a general
discussion on the behaviour of silicates.

The largest quantities of a given dust species is determined by the surface mass fraction of
the so called key-species, defined as the least abundant among the chemical elements 
involved in the formation reaction \citep{fg06}. This is silicon for silicon carbide and  silicates, 
aluminium for alumina dust and the excess of carbon with respect to oxygen for amorphous 
carbon.

To calculate the dust produced by the stars during their life, we consider between 20 and 30 
points during each interpulse phase, chosen in a way such that the main physical quantities of the star do not vary meaningfully. We model dust formation on the basis of the
corresponding values of mass, mass loss rate, luminosity, effective temperature and
the surface chemical composition. This procedure, coupled with the known gas mass loss rate,
allows the computation of the dust production rate during different evolutionary
phases. The integration of the current dust production rate over the whole AGB life
allows the calculation of the total dust mass produced by a star during the AGB 
evolution.

\begin{table*}
\caption{The main properties of the stellar models metallicity $Z=10^{-4}$ and
$Z=3\times 10^{-4}$. The 
different columns report the initial mass of the star (1), the duration of the core 
H-burning (2) and of the TP-AGB phases (3), the number of thermal pulses experienced 
during the AGB evolution (4), the maximum luminosity reached during the AGB phase (5), 
the core mass at the beginning of the TP-AGB phase (6), the fraction of the TP-AGB life 
spent in the C-star phase (7), the final surface mass fractions of 
helium (8), carbon (9), nitrogen (10) and oxygen (11).
}
\begin{tabular}{c c c c c c c c c c c c}        
\hline       
$M/M_{\odot}$ & $\tau_H$ (Myr) & $\tau_{\rm AGB}$ (kyr) & NTP & $L_{\rm max} (10^3L_{\odot})$ & 
$M_{\rm core}/M_{\odot}$ & $\%(C_{\rm star})$ & Y & X(C) & X(N) & X(O) & C$/$O  \\
\hline  
& & & & & $Z=10^{-4}$ & & & & & \\
\hline
 1.0  &  5100  & 1000 &   6   & 5.4  & 0.50 &  59 & 0.272 & 5.5e-3  &  1.7e-5  & 3.8e-4  & 19.3   \\
 1.1  &  3600  & 1550 &   8   & 7.5  & 0.54 &  62 & 0.281 & 5.6e-3  &  2.0e-5  & 5.0e-4  & 14.9   \\
 1.25 &  2300  & 1600 &   11  & 9.2  & 0.55 &  70 & 0.285 & 9.7e-3  &  4.3e-5  & 8.8e-4  & 14.7   \\
 1.5  &  1310  & 1700 &   15  & 11   & 0.57 &  59 & 0.289 & 1.5e-2  &  4.7e-5  & 1.3e-3  & 15.4   \\
 2.0  &  575   & 1500 &   29  & 16   & 0.61 &  76 & 0.296 & 1.8e-2  &  4.9e-5  & 2.3e-3  & 10.4   \\
 2.5  &  340   & 650  &   27  & 26   & 0.72 &  61 & 0.275 & 6.2e-3  &  1.2e-2  & 4.7e-3  & 1.8    \\
 3.0  &  230   & 320  &   28  & 29   & 0.81 &  34 & 0.285 & 1.4e-3  &  3.5e-3  & 8.6e-4  & 2.2    \\ 
 3.5  &  167   & 270  &   35  & 38   & 0.83 &  26 & 0.310 & 1.3e-3  &  3.0e-3  & 6.3e-4  & 2.8    \\
 4.0  &  127   & 190  &   41  & 43   & 0.86 &  25 & 0.327 & 1.0e-3  &  2.3e-3  & 4.4e-4  & 3.0    \\
 4.5  &  101   & 203  &   53  & 52   & 0.90 &  18 & 0.343 & 7.9e-4  &  1.5e-3  & 3.6e-4  & 2.9    \\
 5.0  &  82    & 210  &   103 & 95   & 0.94 &  72 & 0.369 & 5.1e-4  &  1.0e-3  & 3.4e-4  & 2.0    \\
 5.5  &  68    & 230  &   112 & 99   & 0.98 &  86 & 0.383 & 4.0e-6  &  1.8e-4  & 1.0e-6  & 5.3    \\    
 6.0  &  58    & 180  &   121 & 120  & 1.04 &  88 & 0.391 & 3.0e-6  &  6.8e-5  & 3.0e-7  & 13.3   \\
\hline
& & & & & $Z=3\times 10^{-4}$ & & & & & \\  
\hline
 1.0  &  5100  & 1300 &   6   & 5.4  & 0.53 &  23 & 0.275 & 1.5e-3  &  3.0e-5  & 2.9e-4  & 6.9  \\
 1.1  &  3600  & 1350 &   6   & 6.7  & 0.54 &  52 & 0.278 & 3.1e-3  &  2.6e-5  & 5.6e-4  & 7.4  \\
 1.3  &  2050  & 1600 &   9   & 7.8  & 0.54 &  56 & 0.279 & 5.3e-3  &  2.5e-5  & 5.6e-4  & 12.6 \\
 1.5  &  1320  & 1350 &  12   & 9.7  & 0.57 &  59 & 0.282 & 1.1e-2  &  3.3e-5  & 1.3e-3  & 11.3 \\
 1.75 &  860   & 1100 &  14   & 12   & 0.60 &  73 & 0.283 & 1.1e-2  &  5.0e-5  & 1.5e-3  & 9.8  \\
 2.0  &  610   & 1700 &  29   & 16   & 0.60 &  77 & 0.290 & 1.2e-2  &  6.4e-5  & 2.2e-3  & 7.3  \\
 2.5  &  360   &  740 &  31   & 26   & 0.72 &  55 & 0.274 & 2.4e-3  &  1.6e-2  & 4.4e-3  & 0.7  \\
 3.0  &  240   &  320 &  28   & 27   & 0.80 &  29 & 0.277 & 1.2e-3  &  3.6e-3  & 8.0e-4  & 2.0  \\
 3.5  &  174   &  262 &  35   & 35   & 0.83 &  15 & 0.307 & 9.0e-4  &  2.9e-3  & 7.3e-4  & 1.6  \\
 4.0  &  132   &  200 &  42   & 44   & 0.86 &   5 & 0.328 & 8.0e-4  &  1.9e-3  & 4.9e-4  & 2.2  \\
 4.5  &  105   &  170 &  47   & 52   & 0.89 &  23 & 0.341 & 1.8e-4  &  1.7e-3  & 1.4e-4  & 1.7  \\
 5.0  &  80    &  123 &  63   & 58   & 0.94 &  50 & 0.353 & 1.1e-4  &  7.2e-4  & 1.1e-4  & 1.3  \\
 5.5  &  69    &   85 &  54   & 72   & 0.99 &  67 & 0.360 & 7.0e-5  &  4.9e-4  & 1.7e-5  & 5.5  \\
 6.0  &  59    &   61 &  46   & 94   & 1.03 &  75 & 0.367 & 3.6e-5  &  2.0e-4  & 7.2e-6  & 6.7  \\
 6.5  &  51    &   54 &  28   & 112  & 1.16 &  43 & 0.367 & 4.0e-6  &  1.9e-4  & 7.8e-7  & 6.8  \\
 7.0  &  44    &   19 &  23   & 135  & 1.22 &  40 & 0.370 & 9.0e-6  &  2.3e-4  & 3.6e-6  & 3.3  \\
 7.5  &  40    &   11 &  19   & 175  & 1.34 &  8  & 0.369 & 1.6e-5  &  2.9e-4  & 6.0e-6  & 3.5  \\
\hline
\label{tabmod}
\end{tabular}
\end{table*}

\begin{figure*}
\begin{minipage}{0.48\textwidth}
\resizebox{1.\hsize}{!}{\includegraphics{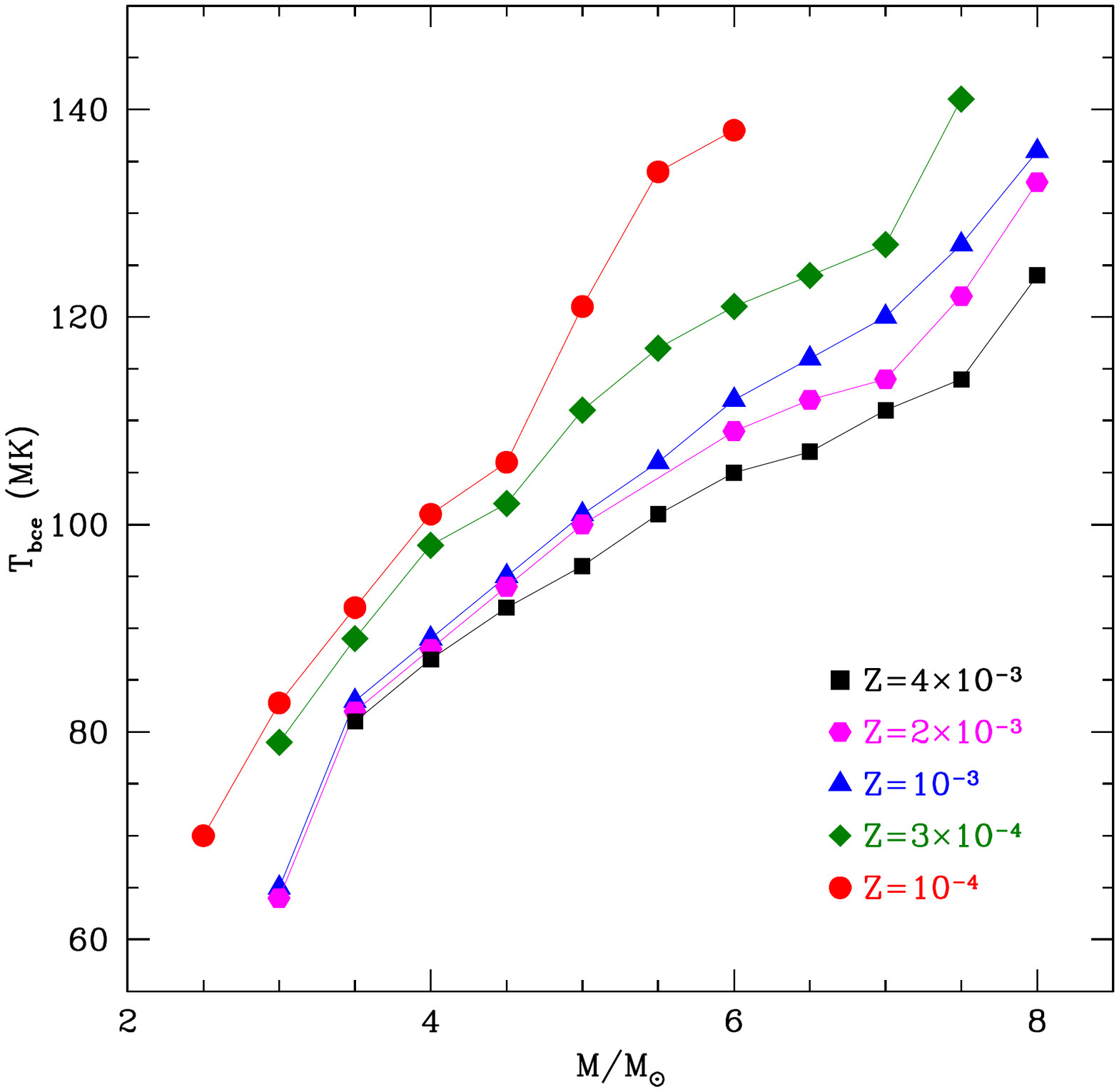}}
\end{minipage}
\begin{minipage}{0.48\textwidth}
\resizebox{1.\hsize}{!}{\includegraphics{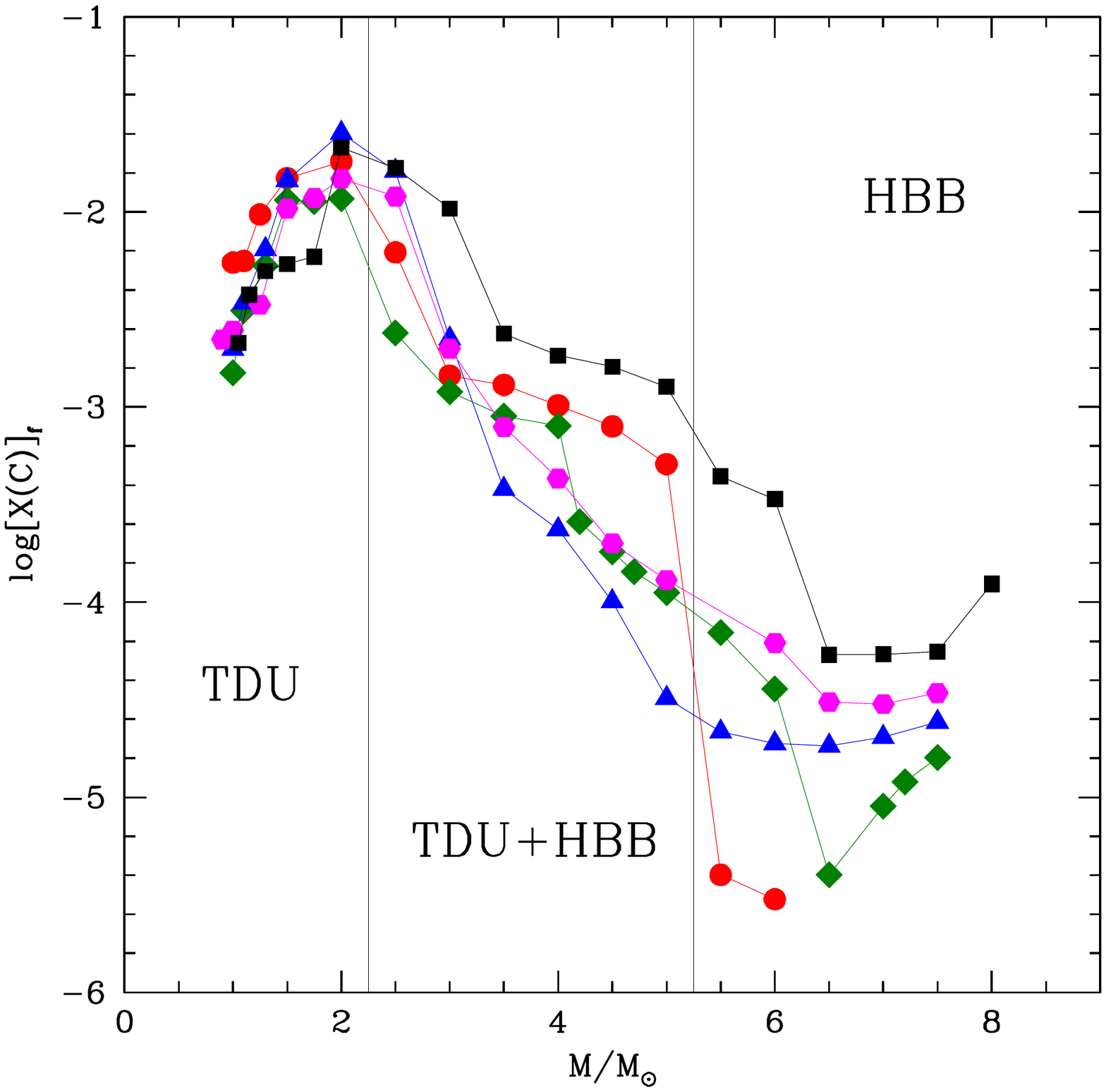}}
\end{minipage}
\vskip-80pt
\begin{minipage}{0.48\textwidth}
\resizebox{1.\hsize}{!}{\includegraphics{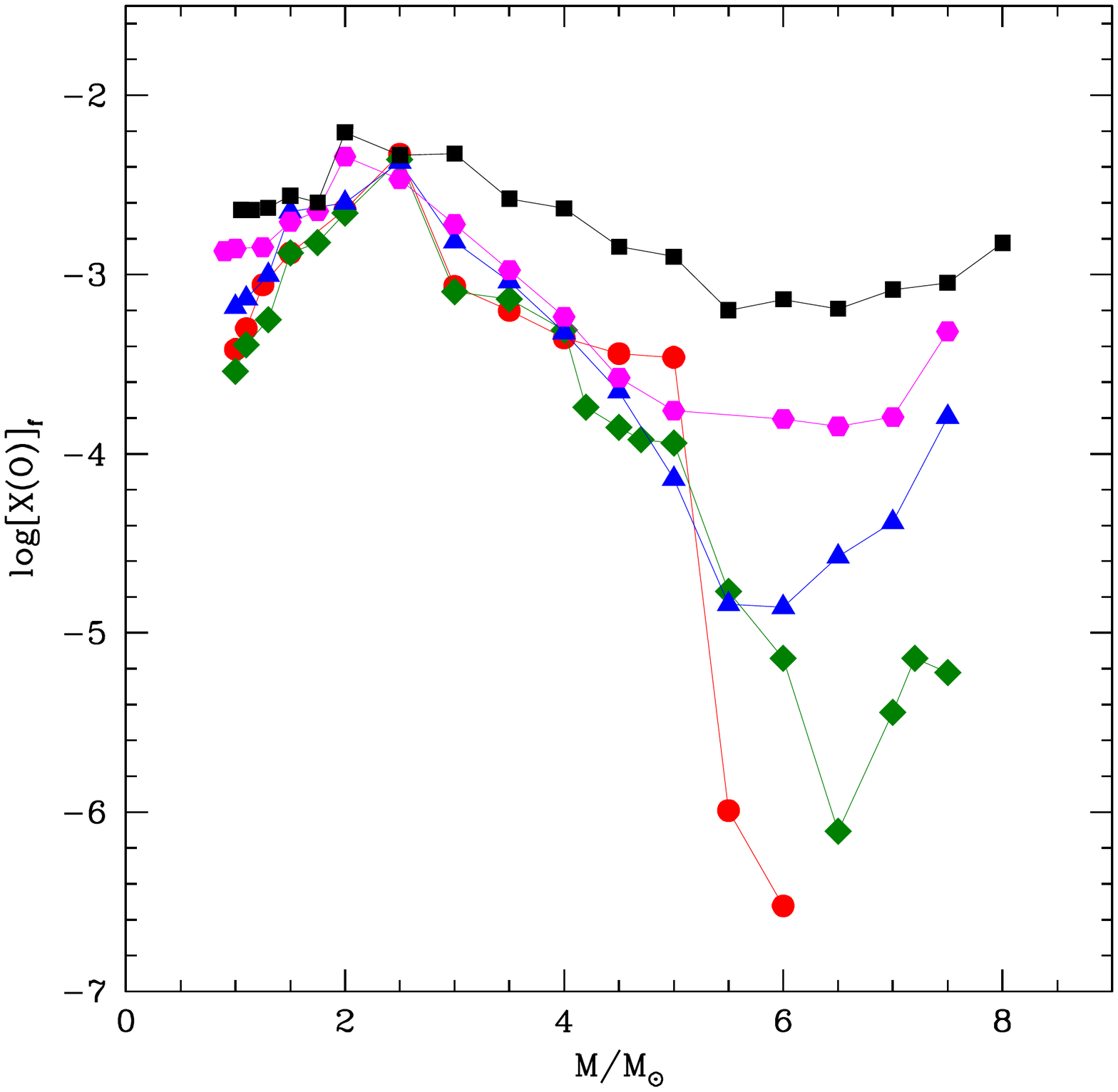}}
\end{minipage}
\begin{minipage}{0.48\textwidth}
\resizebox{1.\hsize}{!}{\includegraphics{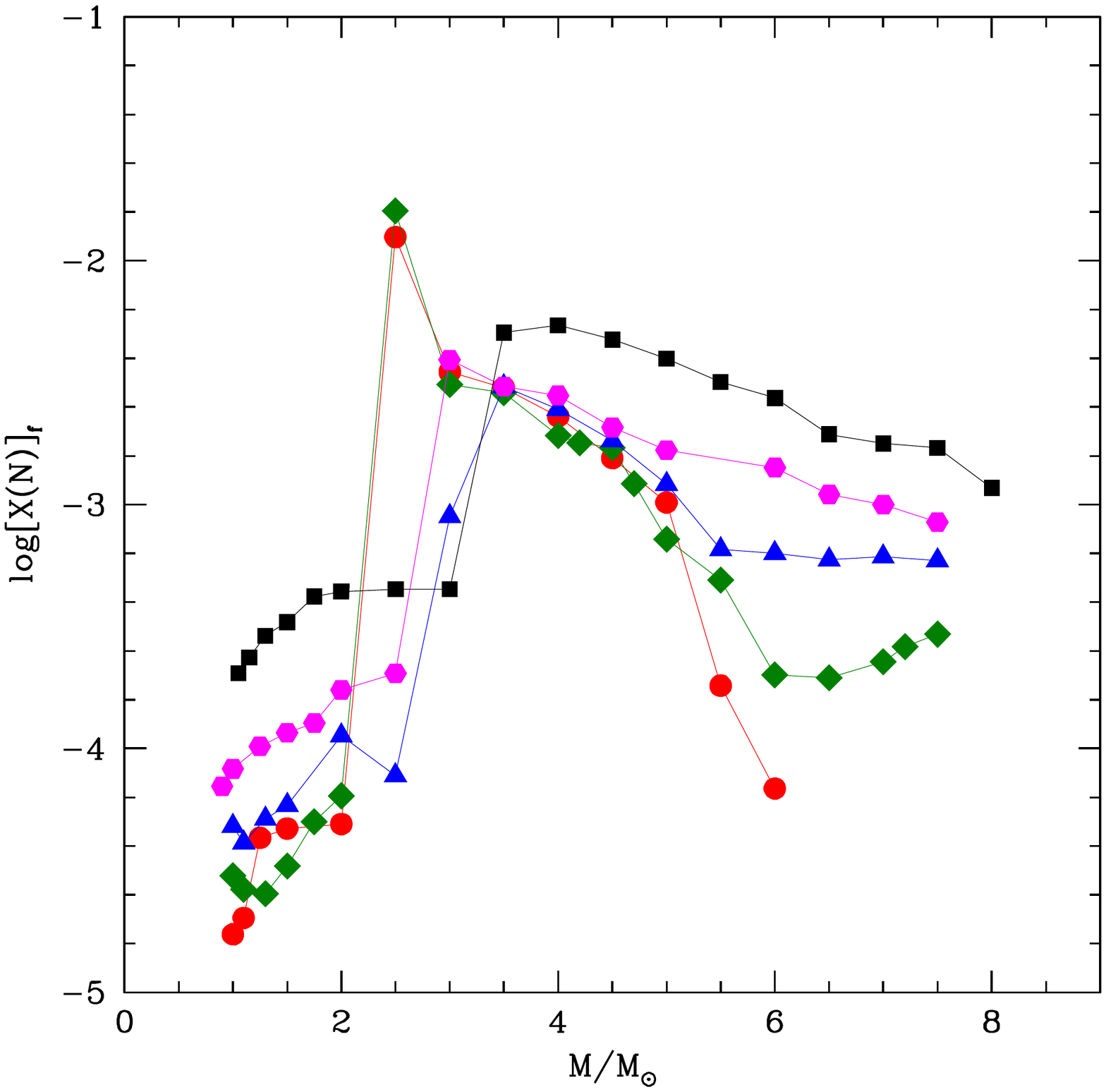}}
\end{minipage}
\vskip-60pt
\caption{The average temperature at the base of the envelope (top, left panel), the final
abundances of carbon (top, right), oxygen (bottom, left) and nitrogen (bottom, right)
are shown as a function of the initial mass. The different metallicities shown are: 
$Z=4\times 10^{-3}$ (black squares), $Z=2\times 10^{-3}$ (magenta exagons), 
$Z= 10^{-3}$ (blue triangles), $Z=3\times 10^{-4}$ (green squares) and $Z=10^{-4}$ 
(red dots).}
\label{fchemf}
\end{figure*}

\section{The AGB evolution of low-metallicity stars}
The most relevant physical phenomena taking place during the AGB evolution are
thoroughly described in the reviews by \citet{herwig05} and \citet{karakas14}, where
the interested reader can find an exhaustive discussion on the mechanisms
potentially able to alter the surface chemical composition, namely third dregde-up
(TDU) and HBB. 

TDU takes place after each thermal pulse, if the 
surface convection, during the inwards penetration, reaches regions previously touched 
by helium burning \citep{iben74, iben75, lattanzio89}: 
the consequence of TDU is the progressive enrichment in carbon 
of the envelope, which eventually favours the conversion of M stars to carbon stars,
with a surface C$/$O above unity. 

HBB occurs in AGB stars evolving on
CO cores of mass higher than $\sim 0.8~M_{\odot}$ and consists into the ignition of
p-capture nuclear activity in the innermost regions of the stellar 
envelope \citep{renzini81, blocker91}; the activation of 
HBB provokes a significant modification of the surface chemical composition, which 
will reflect the equilibria of the p-capture nucleosynthesis occurring at the base of the
external mantle. HBB is experienced by stars of initial mass above $\sim 2.5-3.5~M_{\odot}$,
the threshold value being smaller the lower the metallicity \citep{ventura13}.

In the following we discuss the most important physical and chemical properties of
low-metallicity AGB stars, that can be understood based on the efficiency of TDU and
HBB.

\subsection{The role of mass in the determination of the physical and chemical properties}
\label{evol}
The main properties of the $Z=1,3 \times 10^{-4}$ models calculated for the present
investigation are reported in table \ref{tabmod}. Fig.~\ref{fmodels} shows 
the evolution of the luminosity and of the surface mass fraction of the CNO elements
for some values of the initial mass, chosen to cover the entire range of masses of the 
stars undergoing the AGB evolution. Note that we report the current mass of the star
on the abscissa, which allows to show all the models in the same plane (the different evolutionary
times prevent doing this using the AGB time as abscissa) and, more important,
to have an idea of what is going on during the phase when most of the mass is lost.

\begin{figure*}
\begin{minipage}{0.48\textwidth}
\resizebox{1.\hsize}{!}{\includegraphics{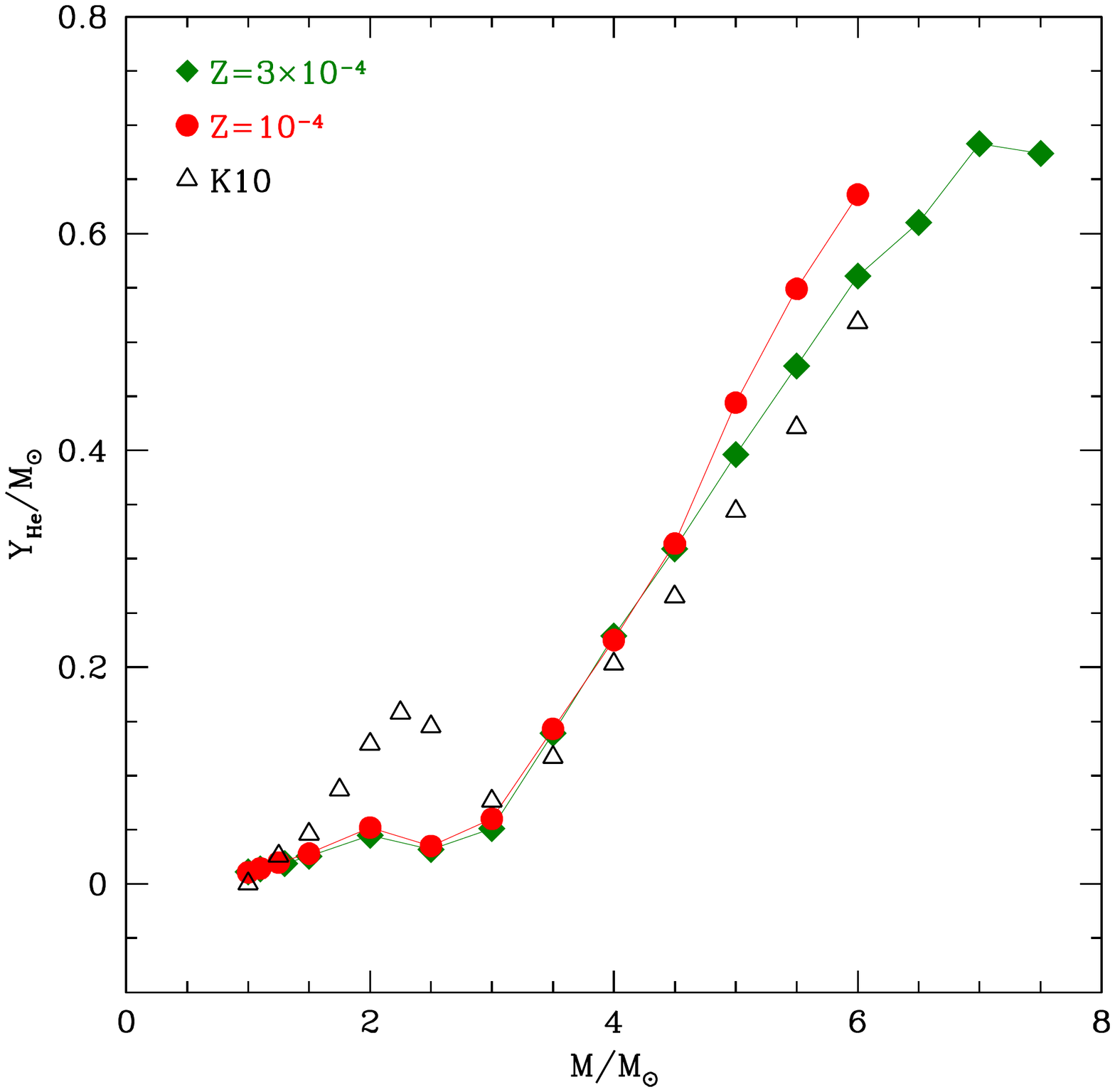}}
\end{minipage}
\begin{minipage}{0.48\textwidth}
\resizebox{1.\hsize}{!}{\includegraphics{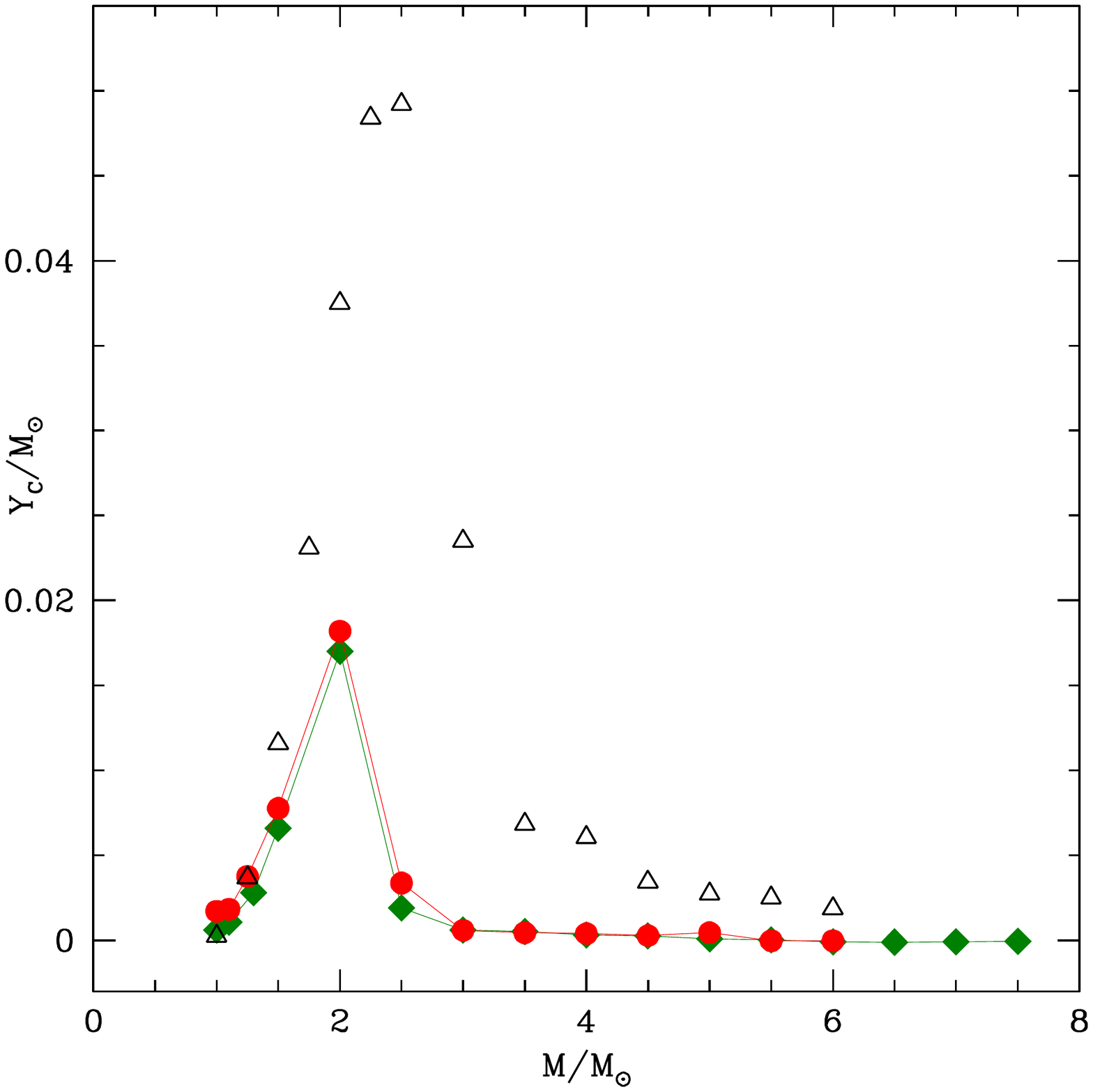}}
\end{minipage}
\vskip-80pt
\begin{minipage}{0.48\textwidth}
\resizebox{1.\hsize}{!}{\includegraphics{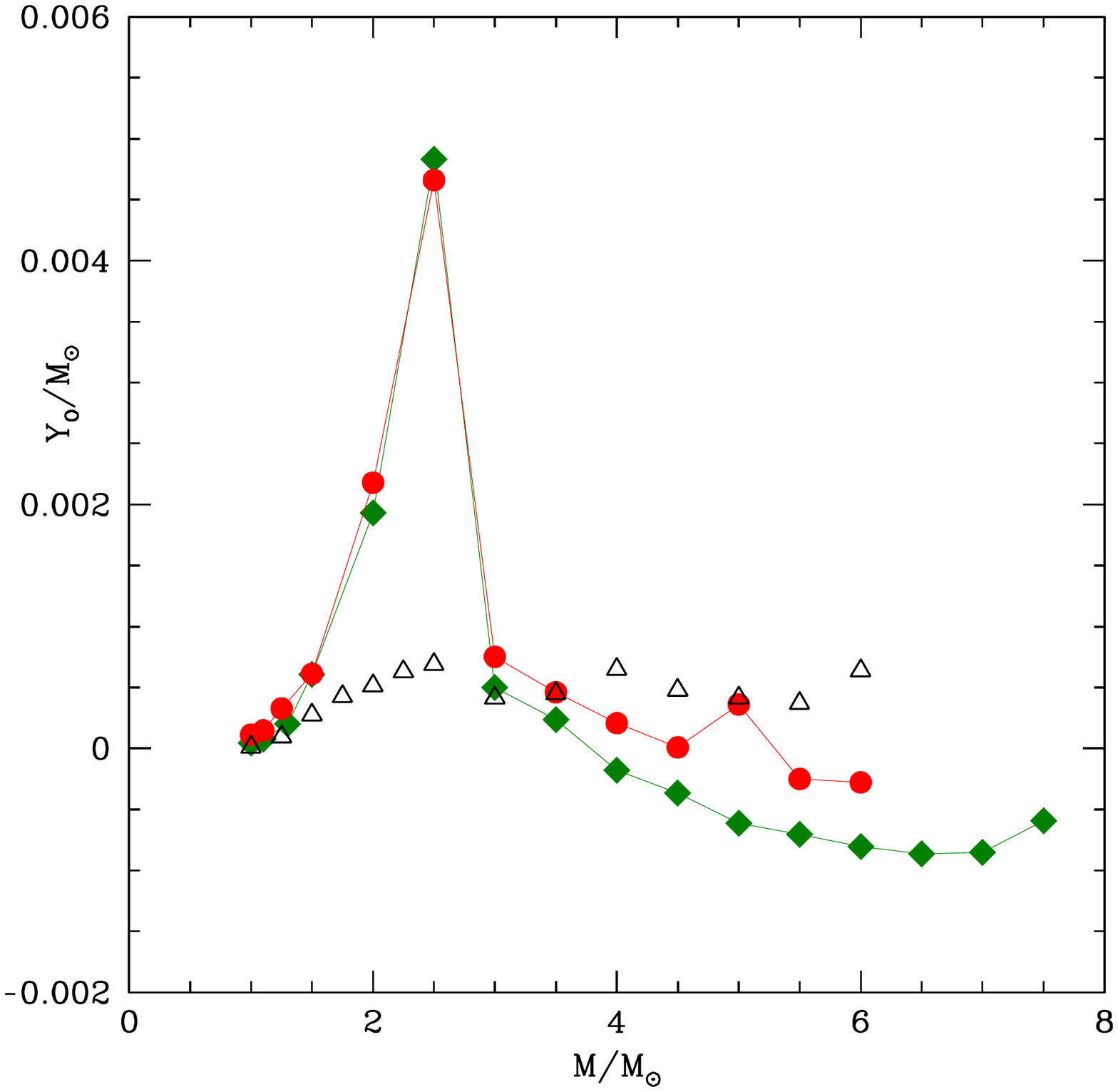}}
\end{minipage}
\begin{minipage}{0.48\textwidth}
\resizebox{1.\hsize}{!}{\includegraphics{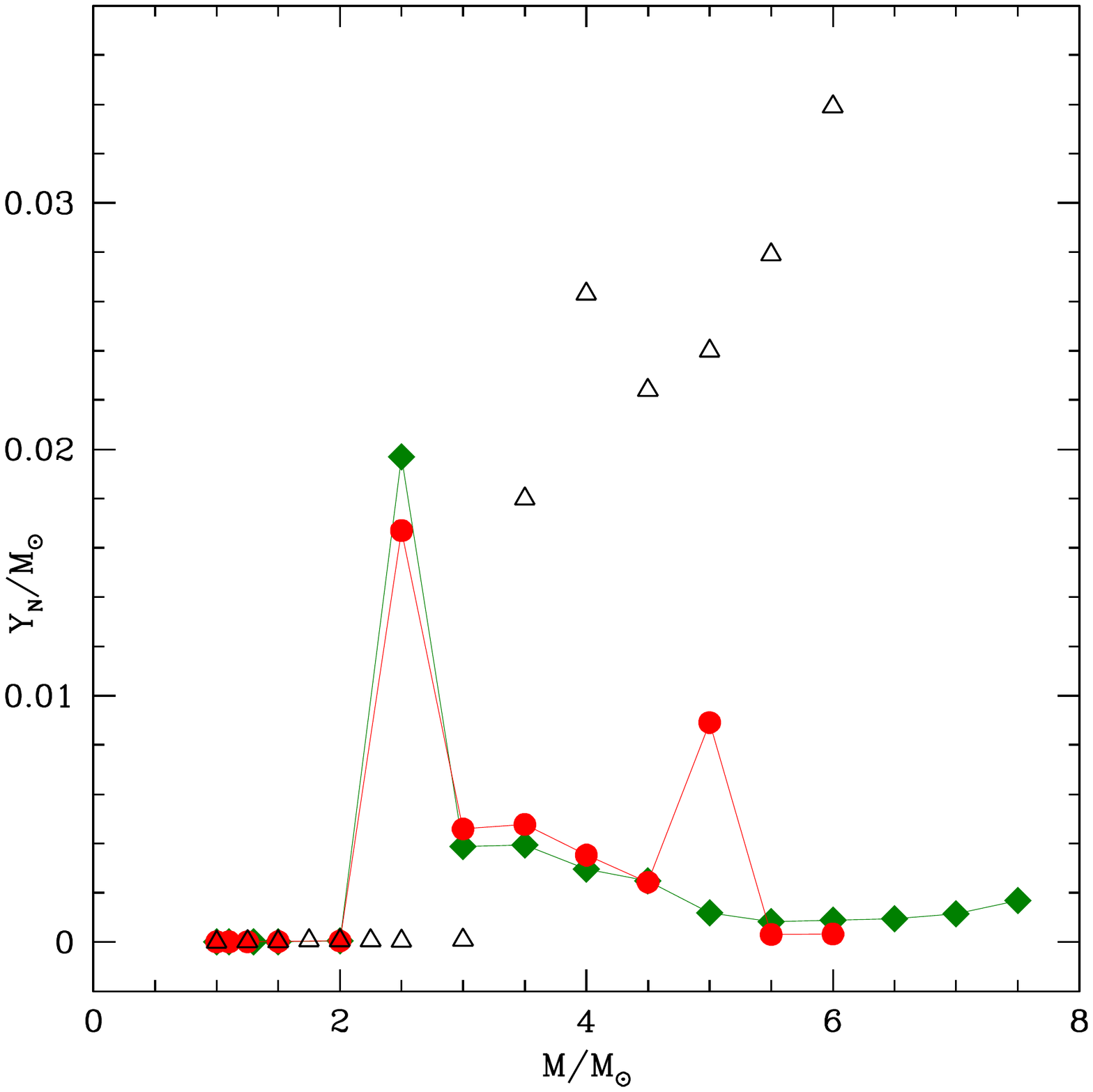}}
\end{minipage}
\vskip-60pt
\caption{Yields of helium and carbon (top panels), oxygen and nitrogen (bottom panels), as a function 
of the initial mass, for $Z=3\times 10^{-4}$ (green diamonds) and $Z=10^{-4}$ (red dots) 
AGB models. The results from \citet{k10}, for $Z=10^{-4}$ models, are indicated with open, 
black triangles. The Y$_i$ values are computed according to Equation \ref{eqyields}.}
\label{fyield}
\end{figure*}

We note in the top, left panel of Fig.~\ref{fmodels} the large sensitivity of the luminosity
of the star to the initial mass. Metallicity is also playing a role here, lower-Z
stars reaching higher luminosities. In low-mass stars (below $\sim 2~M_{\odot}$) the
luminosity increases during the whole AGB phase, owing to the gradual increase in the core
mass. Conversely, in higher mass models, the overall energy release increases during the 
initial AGB phases, then it starts to decline, after a significant fraction of the envelope 
is lost via stellar wind; this well known behaviour of massive AGB stars (see e.g. 
Ventura et al. 2002) is motivated by the decrease in the gravitational energy available, 
which is required to restart CNO burning in the shell after the temporary extinction of 
that nuclear channel, associated to each TP event.

The panels in the right of Fig.~\ref{fmodels} show the main feature regarding the evolution of
low-mass stars, namely the gradual increase in the surface carbon, due to repeated TDU episodes. Once the surface carbon exceeds oxygen, the star becomes a C-star. The panel
in the top reports the variation of the surface carbon mass fraction, whereas in the
bottom, right panel we show $\epsilon($C$-$O$)$, i.e. the carbon excess with respect to
oxygen (in number density), relatively to the number density of hydrogen atoms. The latter quantity
is the key-factor determining the amount of carbonaceous dust that can be formed in
C-rich environments: we will return to this point in the following sections.

In the metal-poor stars discussed here the C-star stage is reached after only a few TPs, 
because the oxygen in the star is very low: the C-star phase is $60\% - 80\%$ of the 
overall AGB evolution for the $Z=10^{-4}$ metallicity, and $20\% - 60\%$ for the 
$Z=3\times 10^{-4}$ case (see table 1). 

The final carbon abundance and $\epsilon($C$-$O$)$ increase with the initial mass of the 
star, because the higher the mass 
the higher the number of TPs experienced before the envelope is lost. 
The amount of carbon accumulated in the envelope is not significantly sensitive
to the metallicity because the carbon convected to the surface is of primary origin,
being produced by $3\alpha$ nucleosynthesis activated in the He-rich buffer at the
ignition of each TP. The gradual increase 
in the surface carbon has important consequences on the evolution of C-stars: the formation of 
CN molecules determines an increase in the opacity of the external regions \citep{marigo02} 
that favours the expansion of the envelope, a rise in the mass loss rate, 
thus a faster consumption of the whole external mantle \citep{vm09, vm10}. 

The variation of the surface chemical composition of the stars experiencing HBB, which, for 
the low metallicities considered here, occurs for initial masses above $\sim 2.5~M_{\odot}$, 
is shown in the bottom, left panel of Fig.~\ref{fmodels}.
For clarity reasons we show only the mass fractions of N and O for a few selected masses,
which represent: a) stars with core mass close to the threshold limit required to core 
collapse, here represented by the models of initial mass $6~M_{\odot}$ ($Z=10^{-4}$) and 
$6.5~M_{\odot}$ ($Z=3\times 10^{-4}$); b) lower mass stars, which experience a 
weaker HBB, indicated in the figure by the tracks corresponding to the 
$4~M_{\odot}$, $Z=10^{-4}$ and $4.5~M_{\odot}$, $Z=3\times 10^{-4}$ models.

The chemical evolution of massive AGBs (case a above) is mostly determined by HBB,
which causes a depletion in the oxygen content and
a parallel increase in the surface nitrogen. In the $Z=10^{-4}$ case oxygen is severely
destroyed in the envelope since the very first TPs, thus the stars reach
very soon the $C/O > 1$ condition, evolving as C-stars for the majority of the AGB life.

In stars of lower mass (case b) the surface 
chemistry is determined by the balance between HBB and TDU; the effects of the
latter mechanism can be seen in
the sudden increase in the surface oxygen abundance, which takes place after
each TP. In these stars the strength of HBB is lower than in their higher mass 
counterparts: the C-star stage is experienced in the initial AGB phases, before the
ignition of HBB favours the destruction of the surface carbon, and in the very final
evolutionary stages, when HBB is turned off, and some carbon is dredged-up to the surface.

All the stars experiencing HBB manufacture large quantities of nitrogen; however,
the final N is higher in lower mass stars, because when TDU is effective, not only the 
initial carbon in the star is available to be turned into N via HBB, but also the
carbon convected to the surface via TDU.

\subsection{The evolution of the surface chemistry in metal-poor AGB stars}
\label{surfchem}
The relative importance of TDU and HBB is sensitive to the metallicity of the star, 
which affects the internal temperatures, the thermodynamical stratification of the external 
mantle and the initial chemical composition.

Regarding the strength of HBB, we show in the top, left panel of Fig.~\ref{fchemf} the
values of the temperature at the base of the envelope, $T_{\rm bce}$, of stars of 
different mass and metallicity\footnote{The temperature at the base of the envelope is
not constant during the AGB phase. The quantities shown here refer to the phase during
which most of mass loss occurs, during which $T_{\rm bce}$ is approximately constant.};
only the stars which experience HBB are indicated. These results confirm previous
findings, that the strength of HBB is higher the lower 
is Z \citep{ventura13}.

To understand the changes in the surface chemical composition occurring during the
AGB evolution, we show in the other three panels of Fig.~\ref{fchemf} the final
surface mass fraction of the CNO species. 

As shown in the top, right panel of Fig.~\ref{fchemf}, low-mass stars produce carbon. 
Consistently with the results shown in the right panels of Fig.~\ref{fmodels}, the trend 
of the final C with 
mass is positive for $M \leq 2~M_{\odot}$, with metallicity playing a minor role in this
context. The largest carbon abundances are of the order of $X(C) \sim 0.02-0.03$. 
Because the final carbon mass fraction is not sensitive to Z, the production factor,
which indicates the ratio between the final and the initial carbon at the surface of
star, changes significantly with the metallicity: in the most metal-poor
cases discussed here, $Z=10^{-4}$, the final carbon is enhanced by a factor up to 
$\sim 2000$, whereas in the $Z=0.004$ models the largest C enrichment is by a factor 
$\sim 40$.

For the stars experiencing HBB the trend of the final surface carbon with mass is negative,
because, as shown in the top, left panel of Fig.~\ref{fchemf}, the higher the initial mass 
the stronger the HBB at the base of the envelope, the faster the carbon destruction process
via proton fusion. In this mass domain the results are sensitive to the metallicity,
because in metal-poor stars the initial, overall C+N+O is smaller and the HBB is more
efficient; both factors lead to a final, lower carbon content in the surface regions
of the star.

Post-AGB stars are the immediate progeny of AGB stars and are unique tracers of the 
nucleosynthesis that occurs prior to and during the AGB phase. Detailed chemical 
abundances studies of four single post-AGB stars in the Magellanic Clouds 
(Van Aarle et al., 2013, De Smedt et al., 2014), with initial masses $1-1.5~M_{\odot}$ 
and $Z = 0.007 - 0.008$, have shown that the observed C$/$O ratio is in the range 
$1.5 - 2.5$. We note that there are currently no observations of post-AGB stars with 
$Z \sim 0.0001$, so we use the existing sample for comparison purposes. The observed C$/$O 
ratios are significantly lower than the predicted C$/$O ratios ($\sim 15$ to 20, see table 1).  
To fully  investigate the discrepancy in the observed and predicted C$/$O ratios and make 
an accurate comparison to the low-metallicity models presented in this study, we require 
a statistically larger set of observations from post-AGB stars that cover a wide range of 
initial masses and probe lower metallicities.

The final carbon mass fractions predicted by the present models of AGB stars were shown 
to bracket the range of values measured in the Planetary Nebulae in the Magellanic 
Clouds \citep{ventura15b, ventura16a}.

Among the CNO species, oxygen is the most sensitive to the metallicity. The trend of the
final O with mass, shown in the bottom, left panel of Fig.~\ref{fchemf}, is qualitatively
similar to C, with low-mass stars showing up some O enrichment and massive AGBs producing
O-poor gas. The O-enrichment associated to TDU is significantly 
smaller than C, with a production factor that is below $\sim 10$ for $Z=0.004$ and
up to $\sim 40$ for $Z=10^{-4}$. The role of metallicity is particularly relevant for the
stars experiencing HBB: AGB stars of metallicity
$Z=0.004$ experience a mild depletion of oxygen, restricted to the most massive objects,
whereas for the lowest metallicities we find a strong depletion of the surface O,
up to a factor $\sim 100$ in $Z=10^{-4}$ stars. The very low oxygen content in the ejecta of
low-metallicity, massive AGB stars was the main argument in support of a possible role
played by this class of objects in the self-enrichment process of globular clusters
\citep{ventura01, dercole08}.

The results regarding the amount of nitrogen produced, shown in the bottom, right panel
of Fig.~\ref{fchemf}, are understood based on the behaviour of carbon and oxygen. Some
N production occurs in low-mass stars, owing to the effects
of the first dredge-up; the N enhancement 
is typically in the range $10-20$, independently of metallicity. The run of N vs mass
exhibits a steep rise in the mass domain close to the lower limit to ignite HBB, owing to
the conversion of C and O to N; this is independent of whether the sole CN or the full 
CNO cycling is activated at the base of the envelope. As discussed earlier in this
section, the AGB stars of mass in the range $3-4~M_{\odot}$ achieve the largest N 
enrichment, owing to the higher availability of carbon. For the massive AGB stars,
in which HBB is the dominant mechanism in changing the surface chemistry, the final N
increases with Z, owing to the higher overall CNO content.

\section{Stellar yields}
The yields of the various chemical species are key quantities to understand the pollution
expected from a class of stars and the way they participate in the gas cycle
of the interstellar medium \citep{kobayashi11}. 

In the following we will use the classic definition, according to which we indicate 
the yield $Y_i$ of the $i$-th element as

\begin{equation}
Y_i=\int{[X_i-X_i^{init}] \dot{M} dt}.
\label{eqyields}
\end{equation}

The integral is calculated over the entire stellar lifetime; $X_i^{init}$ is the mass 
fraction of species $i$ at the beginning of the evolution (references regarding the initial chemical composition are provided in Sec. \ref{inchim}). Based on this definition, the 
yield is negative if an element is destroyed and positive if it is produced over the life 
of the star.

Table \ref{tabyield} reports the yields of different chemical species for the range of
mass and metallicity considered. Fig.~\ref{fyield} shows the yields of helium and of
the CNO elements.

In the top, left panel of Fig.~\ref{fyield} we see that AGB stars of initial mass
above $\sim 3~M_{\odot}$ produce gas enriched in helium. The helium enrichment of the surface
regions occurs during the second dredge-up (SDU) episode \citep{becker79}, which follows 
the end of the
core helium burning phase. The helium yield increases with the initial mass of the star,
reaching a maximum of $\sim 0.7~M_{\odot}$ for the most massive stars experiencing the
TP phase. The helium yields are not significantly sensitive to the metallicity,
as indicated by the almost complete overlapping of the lines corresponding to $Z=10^{-4}$ and
$Z=3\times 10^{-4}$. These results find an explanation in the behaviour of SDU, whose
efficiency, independent of metallicity, increases with the core mass, hence with the
initial mass of the star \citep{ventura10}. In stars of mass below $\sim 3~M_{\odot}$ the 
inwards penetration of the convective envelope, which follows the extinction of helium in 
the core, is not sufficiently deep to penetrate the H-He discontinuity. This is the
reason for the clear discontinuity in the slope of the trend of the helium yield with
mass, present in the two lines in the top, left panel of Fig.~\ref{fyield}.

The yields of carbon, shown in the top, right panel of Fig.~\ref{fyield}, reflects the
evolution of the surface carbon, discussed in section \ref{evol}. Low-mass star produce 
gas enriched in carbon, owing to the effects of repeated TDU events. The largest 
amounts of C are produced by $\sim 2~M_{\odot}$ stars, which attain the largest
surface carbon abundances during the AGB life (see top, right panel of Fig.~\ref{fmodels}):
the carbon yields of these stars are slightly below $0.02~M_{\odot}$. Note that
the carbon yields are almost independent of metallicity in this Z domain, in agreement
with the discussion in section \ref{evol}. 

The carbon yields of stars of mass $M \geq 2.5~M_{\odot}$ are generally close to zero, 
as these stars experience HBB, which destroys the carbon in the envelope. The yields of 
the most massive stars are negative, as the surface chemistry of these objects is mostly
determined by HBB; in stars of mass in the range $2.5~M_{\odot} \leq M \leq 5~M_{\odot}$ 
the carbon yields are positive and negligible, owing to the combined effects of TDU
and HBB.

For what regards oxygen (see bottom, left panel of Fig.~\ref{fyield}), the yields,
similarly to carbon, can be explained based on the balance between the effects of
HBB and TDU. 
Low mass stars produce gas enriched in oxygen, owing to the effects of TDU. The amount 
of oxygen produced (the yield is at most $\sim 0.005~M_{\odot}$, for $2~M_{\odot}$ stars) 
is smaller than carbon. The oxygen yields drop to zero, or turn negative, in the mass domain
$M \geq 2.5~M_{\odot}$, owing to the effects of HBB (see bottom, left panel of Fig.~\ref{fmodels}); 
in the very metal-poor stars discussed here HBB is extremely strong, thus the gas 
produced is oxygen-poor for all the stars experiencing HBB. 

In the $Z=10^{-4}$ case the carbon and oxygen yields of the stars experiencing HBB 
are extremely low, but positive, with the only exception of the $5.5~M_{\odot}$ and $6~M_{\odot}$ stars; 
this behaviour can be explained by considering that in this very low metallicity domain 
a few TDU episodes are sufficient to provoke a significant enrichment in the surface C
and O, which partly counterbalance the effects of HBB.

The effects of TDU, while important in enriching the gas ejected by low-mass AGBs in carbon 
and oxygen, have a scarce influence on nitrogen: the N yields of $M\leq 2~M_{\odot}$ stars 
are below $10^{-4}~M_{\odot}$, showing up a small enrichment, mainly a consequence of the
first dredge-up\footnote{In the present models we did not consider any possibile extra-mixing
from the convective envelope during the red giant phase. In particular, we did not consider
any thermohaline  mixing effects, which might rise the surface N \citep{corinne10}. The N 
yields given here are to be considered as lower limits for the stars with initial mass 
below $\sim 2~M_{\odot}$}. 
The stars exposed to HBB produced N-rich matter due to the
conversion of carbon and oxygen into nitrogen via p-capture nucleosynthesis.
The slope of the N yield with mass is negative in this mass domain because massive AGBs
produce essentially secondary nitrogen, whereas the lower mass counterparts, which experience
several TDU events, produce also large quantities of primary nitrogen. This explains
the peak of $\sim 0.02~M_{\odot}$ at $2.5~M_{\odot}$ in the bottom, right panel of
Fig.~\ref{fyield}.

Regarding the chemical species not involved in CNO cycling, we note in table \ref{tabyield}
that the $^{24}$Mg yields of the stars experiencing HBB are negative, a further 
signature of the high efficiency of HBB in very metal-poor AGB stars. This is consistent 
with the positive yields of aluminium and silicon, which are the products of the
Mg-Al-Si nucleosynthesis \citep{arnould99, ventura11}. Note that the trend of the Al yields 
with mass is not monotonic in the high mass domain because for HBB temperatures
above $\sim 100$ MK the Al equilibrium abundance in the Mg-Al-Si chain decreases.
The sodium yields are either extremely small or negligible in
the large mass domain, because under these conditions the destruction channel of sodium 
prevails over production \citep{mowlavi99}.

\begin{table*}
\setlength{\tabcolsep}{5pt}
\caption{Chemical yields (see Equation \ref{eqyields} for definition) for the AGB models at metallicities $Z=10^{-4}$ and
$Z=3\times 10^{-4}$}.                                       
\begin{tabular}{c c c c c c c c c c c c c c c c c}        
\hline                      
$M$ & $H$ & $He$ & $^{12}C$ & $^{13}C$ & $^{14}N$ & $^{16}O$ & 
$Ne$ & $^{23}Na$ & $^{24}Mg$ & $^{25}Mg$ & $^{26}Mg$ & $^{27}Al$ & $^{28}$Si \\
\hline
& & & & & & $Z=10^{-4}$ & & & & & & &  \\
\hline
1.00 & -1.0E-2 &  1.1E-2 &  1.7E-3 &  1.0E-7 &  5.5E-6 &  1.2E-4 &   0.0E-0  &  1.0E-7 &   0.0E-0 &  8.9E-8 &  9.1E-8 &  3.3E-8 & 0.0E-0  \\
1.10 & -1.6E-2 &  1.4E-2 &  1.8E-3 &  1.1E-7 &  6.1E-6 &  1.5E-4 &   0.0E-0  &  1.2E-7 &   0.0E-0 &  1.2E-7 &  1.4E-7 &  2.9E-8 & 0.0E-0  \\
1.25 & -2.4E-2 &  2.0E-2 &  3.8E-3 &  2.5E-7 &  1.2E-5 &  3.3E-4 &   2.5E-7  &  3.1E-7 &   1.4E-8 &  6.1E-7 &  5.7E-7 &  6.9E-8 & 2.0E-8  \\
1.50 & -3.7E-2 &  2.8E-2 &  7.8E-3 &  4.5E-7 &  1.9E-5 &  6.1E-4 &   1.2E-6  &  1.3E-6 &   1.2E-7 &  5.0E-6 &  3.6E-6 &  6.4E-7 & 8.8E-8  \\
2.00 & -7.3E-2 &  5.2E-2 &  1.8E-2 &  1.1E-6 &  4.1E-5 &  2.2E-3 &   1.3E-5  &  7.9E-6 &   1.6E-6 &  6.1E-5 &  3.1E-5 &  1.7E-5 & 4.0E-7  \\
2.50 & -6.1E-2 &  3.5E-2 &  3.4E-3 &  4.0E-4 &  1.7E-2 &  4.7E-3 &   7.4E-5  &  1.4E-4 &   1.4E-6 &  1.4E-4 &  4.4E-5 &  7.1E-5 & 3.3E-6  \\
3.00 & -6.6E-2 &  6.0E-2 &  5.9E-4 &  5.9E-5 &  4.6E-3 &  7.5E-4 &   7.0E-6  &  3.3E-5 &   4.3E-9 &  9.8E-6 &  3.6E-6 &  4.9E-6 & 4.3E-7  \\
3.50 & -1.5E-1 &  1.4E-1 &  4.5E-4 &  3.9E-5 &  4.8E-3 &  4.6E-4 &   1.4E-5  &  3.1E-5 &  -4.7E-6 &  1.4E-5 &  3.6E-6 &  7.7E-6 & 5.2E-7  \\
4.00 & -2.3E-1 &  2.2E-1 &  3.9E-4 &  4.5E-5 &  3.5E-3 &  2.1E-4 &   3.0E-5  &  2.7E-5 &  -1.2E-5 &  2.0E-5 &  4.1E-6 &  1.4E-5 & 1.0E-6  \\
4.50 & -3.2E-1 &  3.1E-1 &  2.8E-4 &  2.5E-5 &  2.4E-3 &  9.3E-6 &   2.4E-5  &  9.9E-6 &  -1.6E-5 &  1.3E-5 &  1.5E-6 &  1.6E-5 & 4.6E-6  \\
5.00 & -4.5E-1 &  4.4E-1 &  4.5E-4 &  1.0E-4 &  8.9E-3 &  3.6E-4 &   1.6E-4  &  3.9E-5 &  -1.8E-5 &  8.5E-5 &  1.9E-5 &  6.9E-5 & 4.0E-5  \\
5.50 & -5.5E-1 &  5.5E-1 & -2.6E-4 &  3.1E-6 &  3.0E-4 & -2.5E-4 &  -1.2E-5  & -2.1E-7 &  -2.1E-5 & -5.2E-7 & -1.7E-6 & -3.7E-7 & 4.9E-5  \\
6.00 & -6.4E-1 &  6.4E-1 & -2.8E-5 &  3.4E-6 &  3.2E-4 & -2.8E-4 &  -1.3E-5  & -2.9E-7 &  -2.4E-5 & -2.0E-7 & -1.9E-6 & -3.5E-7 & 5.3E-5  \\
\hline 
& & & & & & $Z=3\times 10^{-4}$ & & & & & & &  \\
\hline
1.00 & -1.2E-2 &  1.1E-2 &  5.9E-4 &  8.7E-8 &  1.0E-5 &  4.6E-5 &   -4.8E-8 &  1.4E-7 & -2.9E-9  & -2.4E-10&  1.2E-8 &  5.6E-8 & 0.0E-0  \\
1.10 & -1.5E-2 &  1.4E-2 &  1.1E-3 &  1.1E-7 &  9.2E-6 &  7.7E-5 &   -6.2E-8 &  4.5E-6 &  3.0E-9  &  1.7E-8 &  5.5E-8 &  6.2E-8 & 0.0E-0  \\
1.30 & -2.2E-2 &  1.9E-2 &  2.8E-3 &  2.1E-7 &  1.0E-5 &  2.0E-4 &   -1.0E-8 &  4.4E-7 & -4.4E-9  &  3.4E-7 &  3.6E-7 &  1.4E-7 & 0.0E-0  \\
1.50 & -3.3E-2 &  2.5E-2 &  6.6E-3 &  3.1E-7 &  1.5E-5 &  6.1E-4 &    5.9E-7 &  7.0E-7 & -3.9E-8  &  2.3E-6 &  1.9E-6 &  4.6E-7 & 6.1E-8  \\
2.00 & -6.4E-2 &  4.5E-2 &  1.7E-2 &  1.5E-6 &  5.4E-5 &  1.9E-3 &    1.1E-5 &  6.7E-6 &  8.3E-7  &  5.0E-5 &  2.5E-5 &  1.4E-5 & 3.8E-7  \\
2.50 & -6.1E-2 &  3.2E-2 &  1.9E-3 &  3.1E-4 &  2.0E-2 &  4.8E-3 &    8.2E-4 &  1.6E-4 &  8.9E-4  &  1.2E-5 &  2.9E-4 &  1.4E-4 & 3.0E-5  \\      
3.00 & -5.6E-2 &  5.1E-2 &  5.9E-4 &  7.1E-5 &  3.9E-3 &  5.0E-4 &    2.3E-6 &  2.0E-5 & -4.0E-7  &  5.4E-6 &  1.8E-6 &  2.5E-6 & 6.3E-7  \\
3.50 & -1.4E-1 &  1.4E-1 &  5.2E-4 &  6.2E-5 &  3.9E-3 &  2.4E-4 &    3.9E-6 &  1.8E-5 & -1.1E-5  &  1.4E-5 &  1.3E-6 &  4.8E-6 & 6.9E-7  \\
4.00 & -2.3E-1 &  2.3E-1 &  3.2E-4 &  3.0E-5 &  3.0E-3 & -1.8E-4 &    9.1E-6 &  6.8E-6 & -3.8E-5  &  2.7E-5 &  6.7E-7 &  1.9E-5 & 2.0E-5  \\
4.50 & -3.1E-1 &  3.1E-1 &  2.6E-4 &  3.0E-5 &  2.5E-3 & -3.7E-4 &    2.0E-5 &  2.8E-6 & -5.0E-5  &  2.1E-5 & -8.2E-7 &  3.1E-5 & 9.4E-6  \\
5.00 & -4.0E-1 &  4.0E-1 &  7.4E-5 &  1.6E-5 &  1.2E-3 & -6.1E-4 &    6.3E-6 &  6.6E-7 & -5.8E-5  &  1.1E-5 & -3.8E-6 &  2.1E-5 & 4.1E-5  \\
5.50 & -4.8E-1 &  4.8E-1 &  2.0E-5 &  4.4E-5 &  8.3E-4 & -7.1E-4 &    6.7E-6 &  8.3E-8 & -6.6E-5  &  1.6E-5 & -4.7E-6 &  1.6E-5 & 5.0E-5  \\
6.00 & -5.6E-1 &  5.6E-1 & -8.6E-5 &  7.8E-6 &  8.8E-4 & -8.1E-4 &    7.9E-6 & -1.5E-7 & -7.3E-5  &  2.9E-5 & -5.1E-6 &  1.6E-5 & 4.5E-5  \\
6.50 & -6.1E-1 &  6.1E-1 & -1.0E-4 &  8.4E-6 &  9.4E-4 & -8.6E-4 &    8.4E-6 & -2.0E-7 & -7.9E-5  &  3.5E-5 & -5.4E-6 &  1.8E-5 & 4.3E-5  \\  
7.00 & -6.8E-1 &  6.8E-1 & -1.0E-4 &  1.3E-5 &  1.1E-3 & -8.5E-4 &    1.2E-5 &  3.4E-7 & -8.4E-5  &  5.5E-5 & -5.4E-6 &  1.7E-5 & 2.9E-5  \\
7.50 & -6.7E-1 &  6.7E-1 & -6.9E-5 &  2.1E-5 &  1.7E-3 & -5.9E-4 &    3.8E-4 &  9.7E-6 & -8.1E-5  &  6.9E-5 & -3.8E-6 &  1.0E-5 & 1.2E-5  \\
\hline
\label{tabyield}
\end{tabular}
\end{table*}

\begin{figure*}
\begin{minipage}{0.48\textwidth}
\resizebox{1.\hsize}{!}{\includegraphics{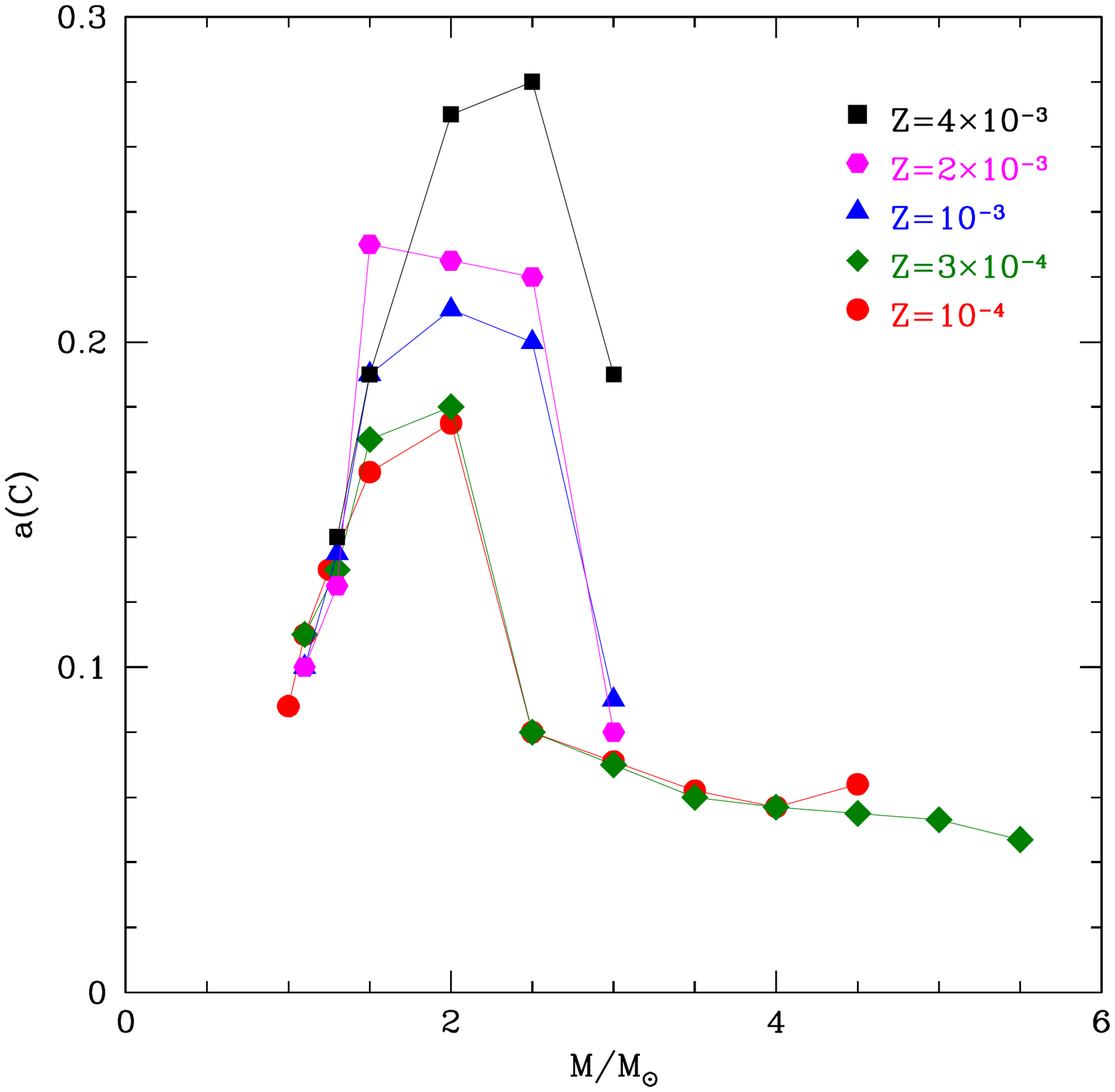}}
\end{minipage}
\begin{minipage}{0.48\textwidth}
\resizebox{1.\hsize}{!}{\includegraphics{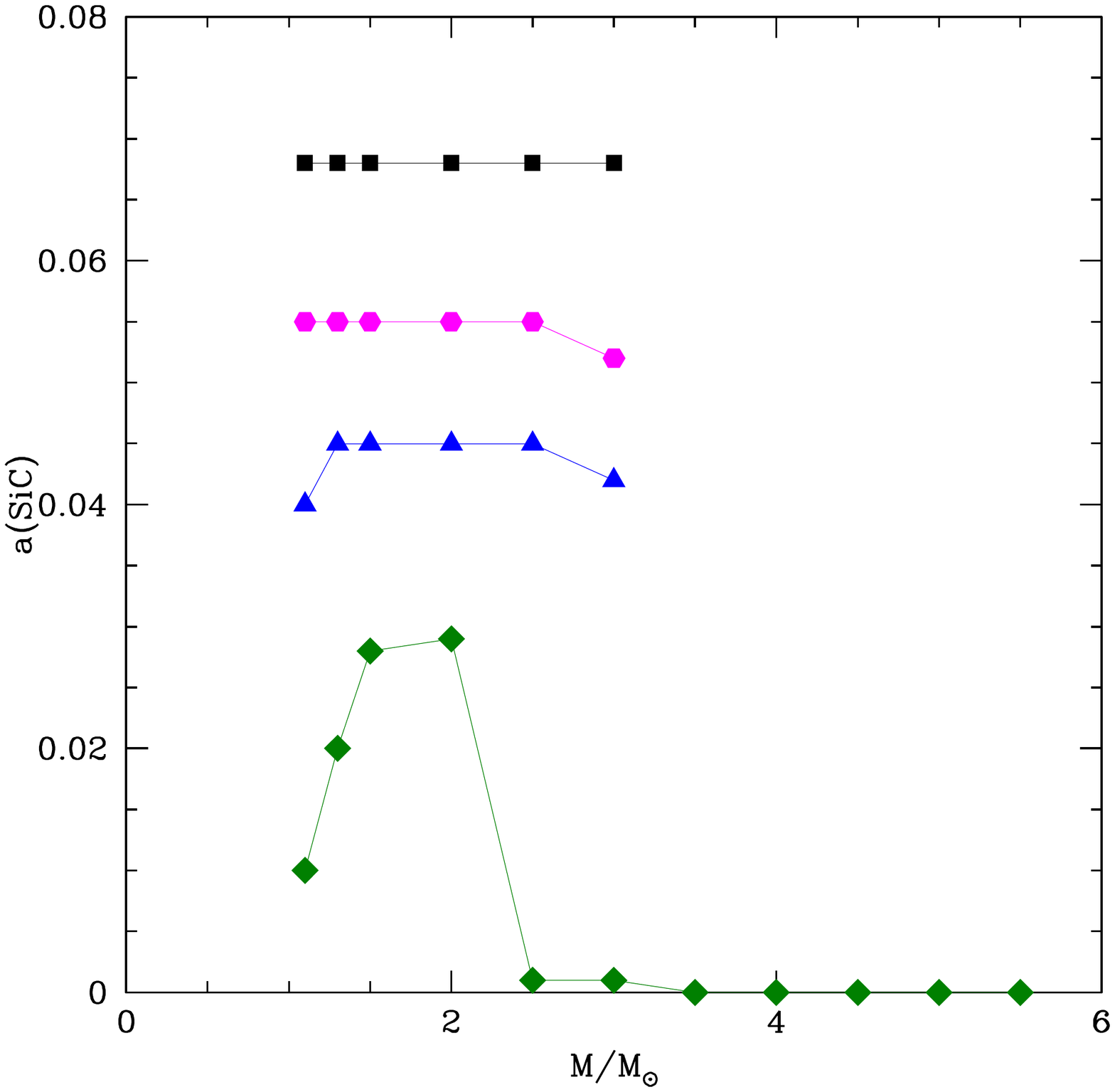}}
\end{minipage}
\vskip-80pt
\begin{minipage}{0.48\textwidth}
\resizebox{1.\hsize}{!}{\includegraphics{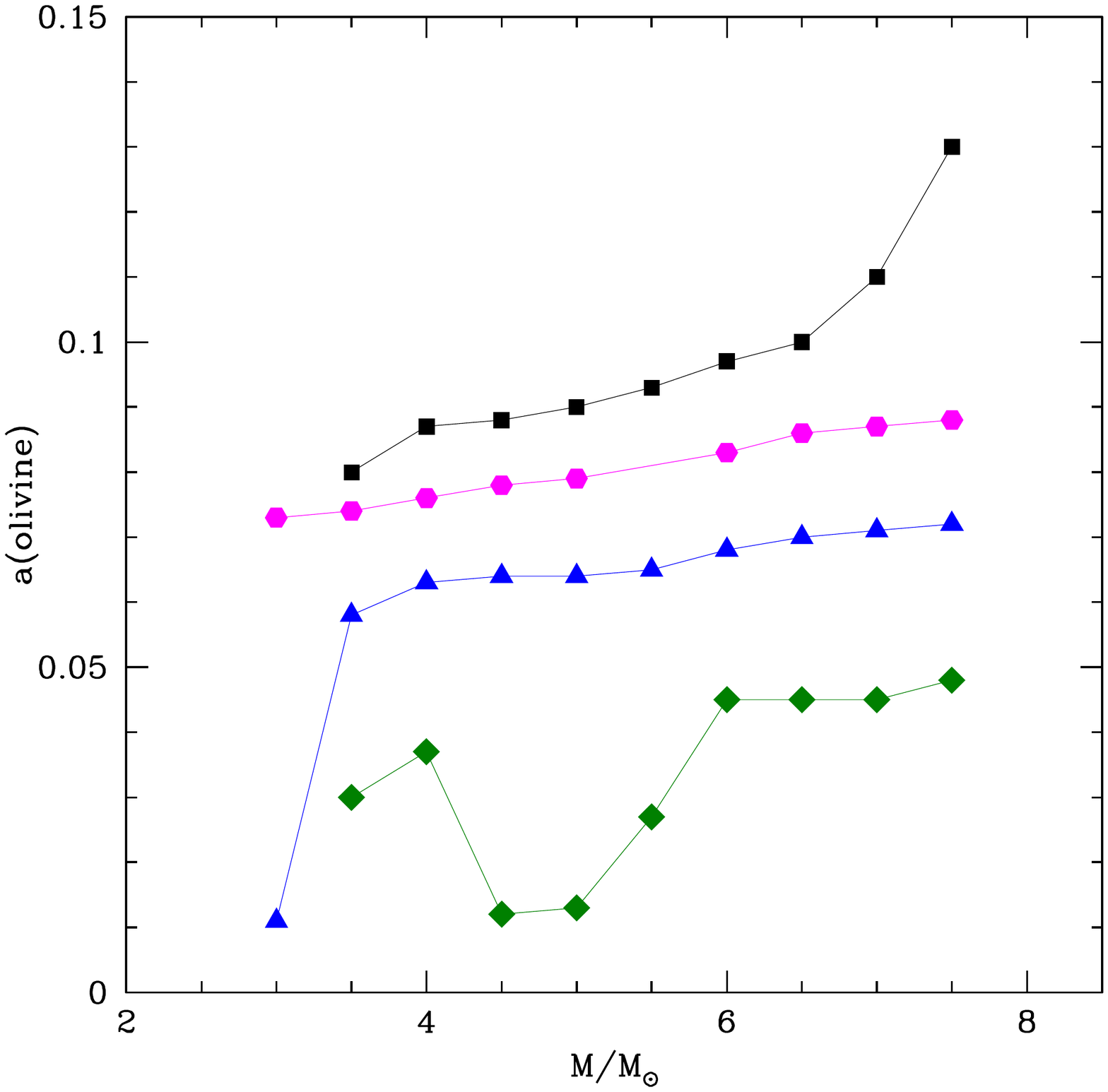}}
\end{minipage}
\begin{minipage}{0.48\textwidth}
\resizebox{1.\hsize}{!}{\includegraphics{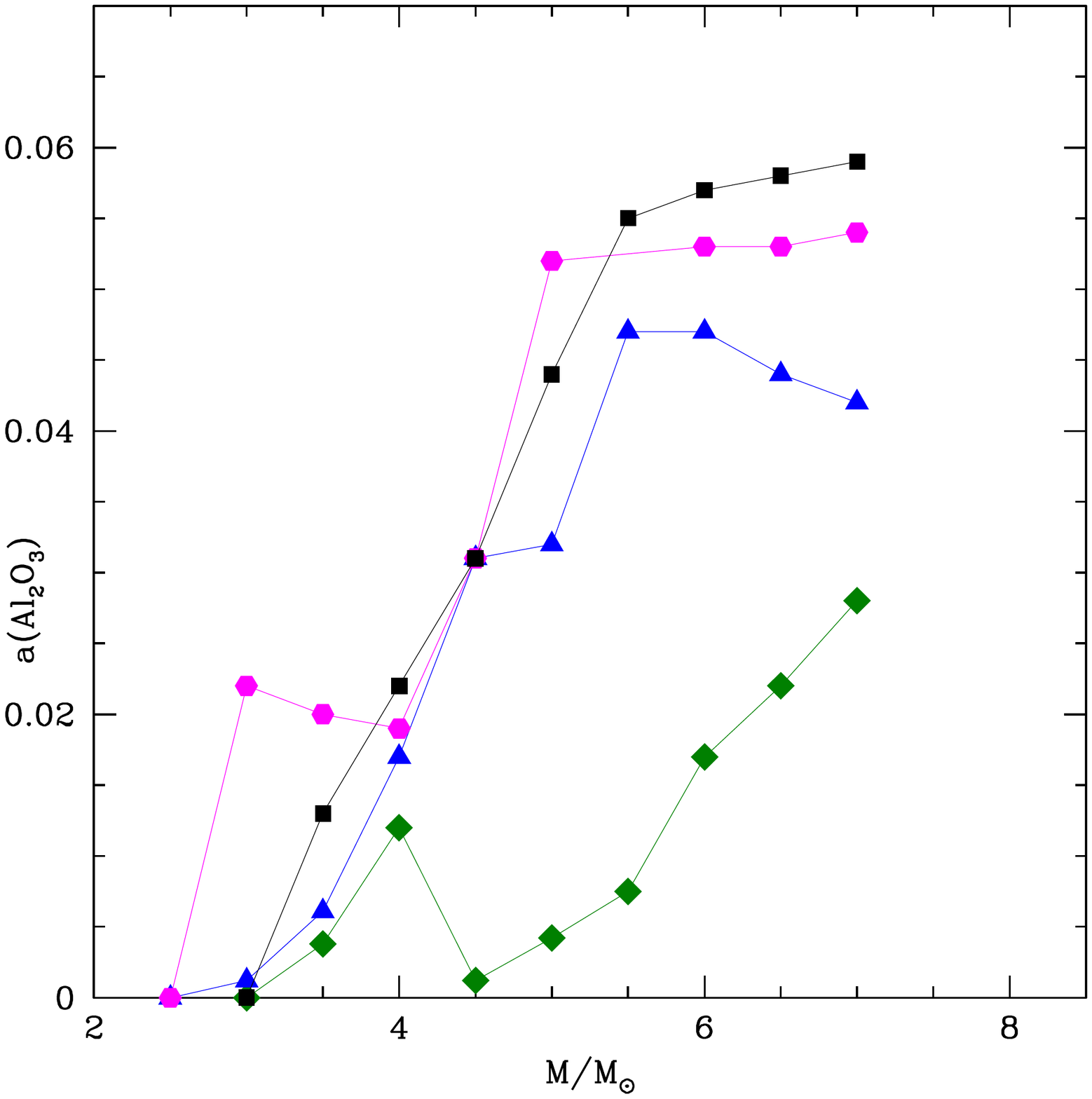}}
\end{minipage}
\vskip-50pt
\caption{The size of the dust grains ($\mu$m units) formed in the winds of AGB stars, 
shown as a function of the initial mass and for different metallicities. Top panels: the 
dust species shown are solid carbon (top-left)and silicon carbide (top-right); the grain sizes shown refers to the end of and the whole AGB phase, respectively. Bottom panels: olivine 
(bottom-left) and alumina dust (bottom-right) grain sizes refer to the phases during which most of the mass loss occurs. For the metallicity we used the same 
symbols as in Fig.~\ref{fchemf}.}
\label{fdust}
\end{figure*}

\section{The comparison with results in the literature}
AGB modelling is affected by  several uncertainties in the 
input macro-phyics, mainly related to convection and mass loss. These uncertainties have 
important effects on the description of the evolution of both low-mass AGBs and of their 
higher mass counterparts, for what regards the main structural and evolutionary properties 
and the alteration of the surface chemical composition.

In the low-mass domain a relevant role is played by the treatment of the convective borders,
particularly of the regions close to the base of the convective envelope; this 
determines the efficiency of TDU and consequently
the extent of carbon and oxygen enrichment occurring after each TP \citep{straniero97}. 
For stars experiencing HBB the key issue is the treatment of the convective instability 
itself, i.e. the modality of calculating the temperature gradient with regions unstable 
to convective motions. The strength of HBB is heavily influenced by the convective
model adopted \citep{vd05}. 

We expect that the impact of these phenomena is particularly relevant for the very
low-Z models discussed in the present investigation. In low-mass stars, because the initial
carbon content of the star is extremely small, a given amount of carbon dredged-up to 
the surface triggers a very large percentage increase in the surface C, which, in turn,
favours a faster and easier achievement of the C-star stage; therefore, the description of TDU 
deeply affects the carbon enrichment in the gas ejected. Regarding stars 
of initial mass above $2~M_{\odot}$, in very low-metallicity AGB stars the efficiency of 
HBB, hence the degree of nucleosynthesis experienced at the base of the external envelope,
changes dramatically according to convection modelling. This is going to affect the
yields of all the species involved in p-capture nucleosynthesis, from carbon to 
silicon.

To understand how the results obtained depend on the description of convection, we report
in Fig.~\ref{fyield} the yields of $Z=10^{-4}$ models, published in Karakas (2010,
hereinafter K10). 

The top-left panel shows that the massive AGB stars helium yields presented here are similar to those by K10. 
This is because the helium enrichment in massive AGB stars
occurs mainly during the SDU, a process, taking place before the beginning of the AGB
phase, whose description is unaffected by the uncertainties associated to the AGB
evolution. This finding confirms the robustness of the helium yields by masssive AGB stars,
in agreement with the analysis by \citet{ventura10}. On the other hand, deeper TDU episodes are responsible for the higher helium yields reported by K10 in the low-mass regime (1.5-3$M_{\odot}$).

Regarding the CNO species, the results shown in Fig.~\ref{fyield} indicate some
similarities and significant differences.

The carbon K10 yields are generally higher than ours. In the low-mass domain this
is due to the higher efficiency of TDU in the K10 computations, which favours
a larger C enrichment. For the stars experiencing HBB, an additional
reason for the differences found is the higher strength of the HBB
in the present models compared to K10; while the K10 carbon yields are positive for
all the masses investigated, in the present case the C yields of massive AGB stars 
are negative, because carbon is severely destroyed by p-capture at the bottom of the
envelope.

The differences in the efficiency of the TDU experienced have important effects on the 
amount of nitrogen produced by AGB stars of mass above $2~M_{\odot}$. 
The N yields are significantly higher in the K10 case, because of the additional 
contribution of primary nitrogen, synthesized via HBB by proton capture on
the carbon nuclei dredged-up during each TDU event.

The strength of HBB has a dominant role in the determination of the oxygen yields.
While the K10 oxygen yields are positive for all the masses considered, in the models 
discussed here the O yields are negative in the large mass domain, owing to the
higher efficiency of the HBB experienced, in comparison to K10.

\section{Dust from metal-poor AGB stars}
The circumstellar envelopes of AGB stars are generally considered as favourable
environments to dust formation \citep{gail99, fg06}. This is mainly due to two reasons: 
a) the effective temperatures of AGB stars are extremely cool, below 5000K, which partly 
inhibits the sublimation process; b) the high number of gas molecules available per unit 
volume, due to the large densities of the wind, triggered by the large rates with which 
these stars lose their envelope (up to a few $\sim 10^{-4}~M_{\odot}/$yr).

As discussed in section \ref{dustinput}, the mineralogy of the dust formed in the wind of 
AGB stars is determined by the C$/$O ratio: carbon stars produce mainly solid carbon and SiC 
particles, whereas oxygen-rich stars produce silicates and alumina dust. In both cases
little amounts of iron dust is formed \citep{ventura12a, ventura12b, ventura14b, flavia17, ventura18}.

Fig.~\ref{fdust} shows the size of the grain particles formed in the wind of
AGB stars of various masses and metallicities. The different panels refer to amorphous
carbon (top, left panel), silicon carbide (top, right), the most stable silicate, i.e. 
olivine (bottom, left), and alumina dust (bottom, right). For what attains low-mass stars, 
we do not consider the formation of silicates before the C-star stage is reached (e.g. 
during the early AGB phases), because in those cases the amount of dust formed is negligible.

We must keep into account that
the size of the solid particles is not constant during the whole AGB evolution, because the
physical conditions within the envelope change and the surface chemical composition is 
altered by the physical processes described earlier in this paper \citep{ventura12a,
ventura12b}. For what attains olivine and alumina dust, the quantities reported in 
Fig.~\ref{fdust} refer to the evolutionary phases during which most of mass loss occurs, 
which are the most relevant to understand the impact of the gas and dust ejected by AGB stars. 
For what regards solid carbon, the dimension of amorphous carbon particles increases
after the C-star stage is reached, owing to the gradual rise in the surface carbon;
the quantities shown in the top, left panel of Fig.~\ref{fdust} refers to the final
AGB phases, during which these stars reach the highest degree of obscuration.
Finally, for what attains SiC, we will see that the species is so highly stable that 
saturation conditions
are easily achieved: the size of the SiC particles keeps approximately constant for the
whole AGB evolution.

The size of the carbon grains formed reflect the behaviour of the surface carbon,
(see top, right panel of Fig.~\ref{fchemf}). For a given metallicity, 
the carbon grains with the largest size are formed in the wind of stars of initial
mass $\sim 2~M_{\odot}$ stars, which in the latest evolutionary phases reach
carbon mass fractions in the envelope above $0.01$. The large degree of obscuration
attained by these stars was proposed by \citet{flavia14, flavia15a} to explain the stars 
with the reddest infrared colours in the LMC.

For metallicities 
$Z \geq 10^{-3}$ carbon particles are produced only in the circumstellar
envelopes of stars of mass below $3~M_{\odot}$, because in their higher mass counterparts
the ignition of HBB prevents the achievement of the C-star condition. In the most 
metal-poor cases some carbon dust is produced even in $M > 3~M_{\odot}$ stars, 
because they become carbon stars as a consequence of the destruction of the surface
oxygen (see discussion in section \ref{evol}); however, in these cases the size of the 
carbon grains is below $\sim 0.05 \mu$m,
because also carbon is exposed to proton fusion under HBB conditions, thus the excess of 
carbon with respect to oxygen in the envelope is very low.
Most of the carbon dust is produced by low-mass stars. 

Interestingly, we see that, according to our modeling, despite the increase in the surface carbon is fairly 
independent of metallicity (see top, right panel of Fig.~\ref{fchemf}), the
size reached by amorphous carbon grains is higher the higher is Z: the grains with
the largest dimension, of the order of $0.28 \mu$m, form in the wind of $Z=0.004$ stars.
This can be explained by considering that metal-poor stars evolve at hotter surface
temperatures, a condition that favours sublimation, which inhibits dust formation.

SiC dust is extremely stable: compared to carbon, SiC grains form closer to the
surface of the stars, $\sim 2$ stellar radii from the photosphere \citep{fg02}. 
The amount of SiC 
formed is determined by the amount of silicon available, thus it is sensitive to the 
metallicity. Because the SiS molecule is extremely stable \citep{sharp90}, only the gaseous silicon 
unlocked in SiS molecules, i.e. $\sim 50\%$, is available to condense into SiC. 
The size of the SiC grains grows with Z: we find that $a($SiC$) \sim 0.07 \mu$m for
$Z=0.004$, $a($SiC$) \sim 0.055 \mu$m for $Z=0.002$, $a($SiC$) \sim 0.045 \mu$m for
$Z=0.001$ and $a($SiC$) < 0.03 \mu$m for $Z=0.0003$. No SiC forms around
$Z=0.0001$ stars. The size of the SiC particles formed is practically independent of mass,
because saturation conditions are easily reached in all the masses considered, halting a further growth of the grains.

The formation of silicates and alumina dust is sensitive to Z, because 
the size that the grains can reach depends on the amount of silicon and aluminium in the 
envelope. The size of olivine grains
increases with metallicity and the mass of the star, because higher mass AGB stars
experiment stronger HBB conditions. The trend of the size of olivine grains with mass
is not monotonic in the $Z=3\times 10^{-4}$ case, because in stars of mass 
$\sim 4.5~M_{\odot}$ the destruction of the surface oxygen provokes a scarcity of water
molecules, which are required to form silicates. In $Z=10^{-4}$ AGB stars negligible
formation of silicates is expected.

On the qualitative side, alumina dust follows the same behaviour of silicates. The size
of the Al$_2$O$_3$ grains is significantly smaller than silicates, because the amount of 
silicon in the envelope is higher than aluminium.

\begin{figure}
\begin{minipage}{0.48\textwidth}
\resizebox{1.\hsize}{!}{\includegraphics{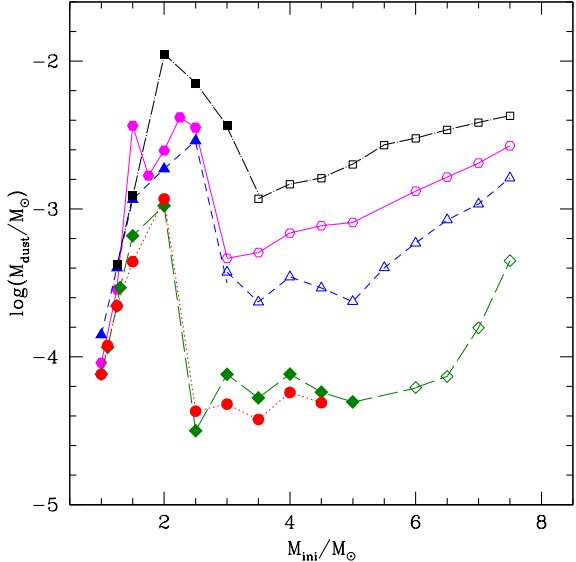}}
\end{minipage}
\vskip-50pt
\caption{Total dust mass produced during the AGB phase for AGB stars of different
initial mass (reported on the abscissa) and metallicities. Color-coding is the same 
as in Fig. \ref{fchemf}. Full points indicate that most of the dust is under the form
of carbonaceous particles, open points correspond to a dominant contribution from
silicates and alumina dust.}
\label{fdtot}
\end{figure}

The summary of the results obtained are reported in Table \ref{tabyielddust}, together with the dust mass produced by low-metallicity stars during the entire AGB phase for each dust species. The total amount of dust is also shown in Fig.~\ref{fdtot}. 
A clear trend, expected based on the arguments discussed earlier in this section, is
the drop in the amount of dust formed by stars of mass above $3~M_{\odot}$ as the
metallicity decreases. The trend of the dust formed with mass also changes for metallicities
$Z < 3\times 10^{-4}$. Indeed we have seen that in  massive AGB stars of metallicity 
$Z = 10^{-4}$ the strong HBB conditions provoke the destruction of the surface 
oxygen and the conversion of stars into C-stars; the surface carbon is extremely
small though, thus dust is produced only in negligible quantities. In this mass domain
the amount of dust produced is below $10^{-4}~M_{\odot}$ for metallicities
$Z < 10^{-3}$.

The only dust produced in meaningful quantities by AGB stars of metallicity $Z < 10^{-3}$
is the amorphous carbon manufactured by $M \leq 2.5~M_{\odot}$, not experiencing HBB.
The mass of solid carbon produced, below $\sim 10^{-3}~M_{\odot}$, is smaller than in their 
higher metallicity counterparts, according to the arguments given above. 

In addition to constraining the nucleosynthesis experienced in the end of the AGB phase, the comparison with the dust that resides in the circumstellar environment surrounding 
single post-AGB stars is also a good tracer of the dust produced in the previous stage
(see Van Winckel 2003 and references therein). Observational studies of the 
circumstellar environment around C-rich single post-AGB stars have shown that 
their dust is predominantly C-rich. For instance, the $11.3 \mu$m amorphous SiC 
feature is commonly observed in the circumstellar environment 
of many of these stars. Additionally, the C-rich post-AGB stars that are also s-process 
enhanced show the $21\mu$m  feature (Kwok et al., 1989), whose exact nature remains unclear 
(Zhang et al., 2009) but is likely to be associated with C-rich dust. In the case of O-rich 
post-AGB stars, silicates (both amorphous and crystalline), oxides, and traces of crystalline 
water ice, make up most of their circumstellar environment (see Waters 2011 and references 
therein). However, we note that the current sample of observed post-AGB stars in the Galaxy 
and Magellanic Clouds are at higher metallicites (Z\,$\sim$\,0.007\,$-$\,0.008, see Section~\ref{surfchem}) than that of 
the models presented in this study. Therefore, to make an accurate comparison to the predicted dust 
chemistry, we require observations from post-AGB stars that 
cover a wide range of initial masses and probe lower metallicities.

\begin{table*}
\setlength{\tabcolsep}{5pt}
\caption{Dust yields produced during the AGB phase for different
initial masses at metallicities $Z=10^{-4}$ and $Z=3\times10^{-4}$. The columns refer to the initial mass of the stars, the total mass of dust and the mass of dust formed for the following species (in order): olivine, pyroxen, quartz, alumina dust, iron, carbon and silicon carbide.}                                       
\begin{tabular}{c c c c c c c c c}        
\hline                      
$M$ & $M_{tot}$ & $M_{ol}$ & $M_{py}$ & $M_{qu}$ & $M_{Al_2O_3}$ & $M_{ir}$ & 
$M_{car}$ & $M_{SiC}$ \\
\hline
& & & & $Z=10^{-4}$ & & & & \\
\hline

1.0 & 7.62D-05 & -- & -- & -- & -- & -- & 7.62D-05 & 7.43D-10\\  
1.1 & 1.19D-04 & -- & -- & -- & -- & -- & 1.19D-04 & 2.84D-09\\
1.25 & 2.21D-04 & -- & -- & -- & -- & -- & 2.21D-04 & 6.05D-09\\
1.5 & 4.40D-04 & -- & -- & -- & -- & -- & 4.40D-04 & 1.78D-08\\
2.0 & 1.17D-03 & -- & -- & -- & -- & -- & 1.17D-03 & 3.15D-07\\
2.5 & 4.28D-05 & 1.44D-07 & 5.80D-08 & 2.22D-08 & 1.75D-04 & 2.17D-08 & 4.26D-05 & 1.22D-09\\
3.0 & 4.78D-05 & 2.10D-07 & 8.70D-08 & 3.53D-08 & 3.51D-06 & 5.35D-08 & 4.74D-05 & 2.03D-07\\
3.5 & 3.77D-05 & 9.19D-07 & 3.61D-07 & 1.42D-07 & 1.26D-05 & 2.16D-07 & 3.61D-05 & 6.39D-07\\
4.0 & 5.73D-05 & 1.00D-06 & 3.95D-07 & 1.58D-07 & 2.08D-05 & 2.67D-07 & 5.54D-05 & 2.21D-06\\
4.5 & 4.89D-05 & 1.08D-07 & 5.90D-08 & 3.71D-08 & 8.80D-06 & 7.05D-07 & 4.80D-05 & 5.79D-06\\
5.0 & 7.28D-07 & 1.27D-09 & 9.71D-10 & 1.01D-10 & 1.08D-09 & 7.25D-07 & -- & --\\
5.5 & 7.92D-07 & 2.87D-09 & 1.58D-09 & 7.91D-10 & 1.63D-09 & 7.85D-07 & -- & --\\
6.0 & 6.01D-07 & 3.31D-08 & 1.66D-08 & 1.03D-08 & 1.56D-07 & 5.41D-07 & -- & --\\
\hline
& & & & $Z=3\times10^{-4}$ & & & & \\
\hline
1.00 & 7.62D-05 & -- & -- & -- & -- & -- & 7.62D-05 & 7.43D-10\\
1.10 & 1.16D-04 & -- & -- & -- & -- & -- & 1.16D-04 & 1.19D-07\\
1.25 & 2.21D-04 & -- & -- & -- & -- & -- & 2.21D-04 & 6.05D-09\\
1.30 & 2.93D-04 & -- & -- & -- & -- & -- & 2.93D-04 & 5.58D-07\\
1.50 & 6.58D-04 & -- & -- & -- & -- & -- & 6.58D-04 & 1.75D-06\\
2.00 & 1.05D-03 & -- & -- & -- & -- & -- & 1.05D-03 & 2.62D-06\\
2.50 & 3.16D-05 & 3.33D-06 & 1.20D-06 & 4.13D-07 & 4.10D-04 & 6.04D-08 & 2.66D-05 & 7.77D-09\\
3.00 & 7.59D-05 & 3.14D-06 & 1.18D-06 & 4.52D-07 & 4.56D-09 & 6.50D-07 & 7.05D-05 & 3.72D-06\\
3.50 & 5.27D-05 & 1.06D-05 & 3.93D-06 & 1.47D-06 & 5.18D-06 & 2.33D-06 & 3.43D-05 & 4.35D-06\\
4.00 & 7.63D-05 & 1.37D-05 & 6.82D-06 & 4.02D-06 & 7.01D-07 & 6.48D-06 & 4.53D-05 & 2.32D-05\\
4.50 & 5.77D-05 & 3.59D-07 & 1.76D-07 & 1.05D-07 & 7.75D-06 & 1.21D-05 & 4.50D-05 & 1.26D-05\\
5.00 & 4.94D-05 & 8.15D-08 & 4.37D-08 & 2.43D-08 & 9.41D-07 & 1.75D-05 & 3.17D-05 & 2.61D-05\\
5.50 & 5.60D-05 & 9.36D-07 & 4.98D-07 & 2.55D-07 & 9.39D-07 & 1.88D-05 & 2.45D-05 & 1.01D-05\\
6.00 & 6.19D-05 & 1.02D-05 & 5.48D-06 & 3.47D-06 & 9.37D-07 & 4.28D-05 & -- & --\\      
6.50 & 7.36D-05 & 1.33D-05 & 4.83D-06 & 2.07D-06 & 6.64D-06 & 5.34D-05 & -- & --\\
7.00 & 1.57D-04 & 5.91D-05 & 3.46D-05 & 9.10D-06 & 2.58D-05 & 5.40D-05 & -- & --\\
7.50 & 4.46D-04 & 3.20D-04 & 9.72D-05 & 2.68D-05 & 3.57D-05 & 2.02D-06 & 2.82D-08 & 3.63D-08\\
\hline
\label{tabyielddust}
\end{tabular}
\end{table*}

\section{The contribution of metal-poor AGB stars to the cosmic dust yield}
\begin{figure}
\centering
\includegraphics[width=9cm]{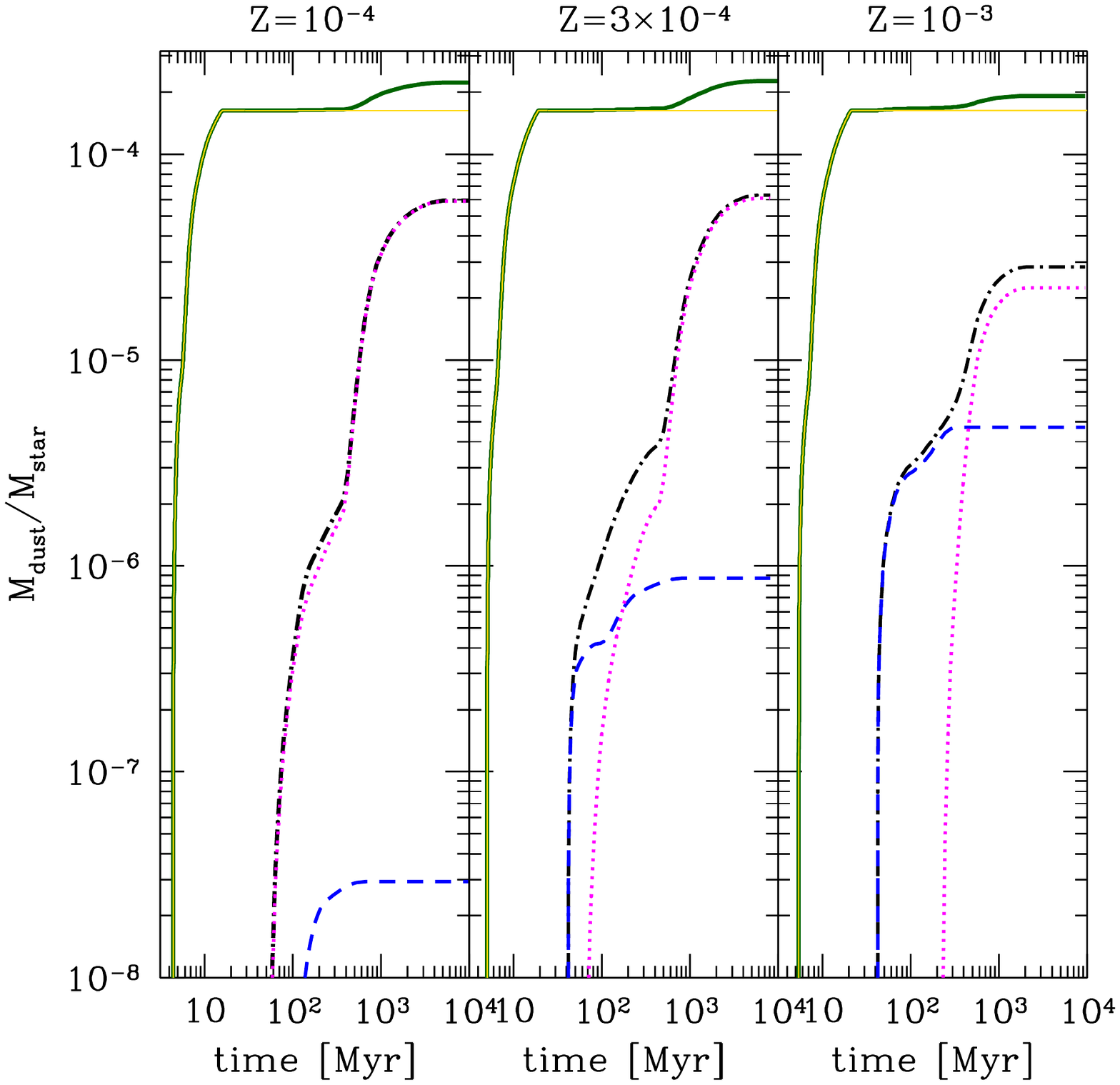}
\caption{Time evolution of the total mass of dust produced by AGB stars per unit stellar 
mass formed in a single burst of star formation for stellar initial metallicity of 
$10^{-4}$ (left-hand panel), $3\times 10^{-4}$ (middle panel), from this work, and 
$Z=10^{-3}$ \citep[right-hand panel; from][]{ventura12b}. At t=0 stars with 
$[0.1-100] \, M_\odot$, at each metallicity, form according to a Larson IMF. Black dot-dashed lines 
show the total dust mass while dotted and dashed curves indicate the contribution of 
carbon and silicate dust, respectively. The green-thick solid lines indicate the cosmic dust 
yield, while the yellow-thin solid lines refer to the contribution of SNe to dust production (see text for details).}
\label{figRosa}
\end{figure}

Following the discussion presented in our previous works, here we 
briefly discuss the impact of the new AGB dust yields, 
computed with ATON, on the so-called cosmic dust yield.

To study the dust contribution from low-metallicity AGB stars to the overall dust
budget of stellar populations we follow the methodology by \citet[][]{valiante09}:
The time evolution of the total dust mass produced by a single population of 
stars with initial mass $[0.1-100] \, M_\odot$ forming at a fixed metallicity 
of $Z=10^{-4}$, $Z=3\times 10^{-4}$ and $10^{-3}$ is computed as:
\begin{equation}
  M_{dust}(t)=\int_{0}^{t} dt' \int_{m(t')}^{100 M_\odot} m_{dust}(m) \phi(m) SFR(t'-\tau) dm
\end{equation}
where $m$ and $m_{dust}(m)$ are the progenitor star and its produced dust mass and SFR 
is the star formation rate that, in general, is a function of time $t'$ and the mass- (and metallicity-) dependent 
evolutionary time-scale, $\tau$.
The lower mass limit, $m(t')$, is set by the stellar mass corresponding to 
an evolutionary time-scale $\tau=t'$, i.e. the minimum stellar mass 
that is able to contribute to dust production at time $t'$. 
The evolutionary time-scales of intermediate mass stars are reported in Table \ref{tabmod}.
The initial mass function (IMF), $\phi(m)$ is a Larson distribution with
characteristic mass $m_{ch}=0.35$ \citep{larson98}\footnote{We normalize to unity the 
integral of $m \phi(m)$ in the mass range $[0.1-100] \, M_\odot$.}:
\begin{equation}
  \phi(m) \propto m^{-\alpha} e^{-m_{ch}/m},
\end{equation}
with $\alpha=2.35$
In this calculation we assume that all the stars form in an instantaneous burst
at $t=0$, integrating the mass- and metallicity-dependent dust injection rate from AGB 
stars and SNe for about 10 Gyr. This simple approach enable us to compare the
maximum contribution of the two main stellar dust sources, AGB and SNe, before 
newly formed grains are injected (and reprocessed) into the host galaxy ISM. 

In Figure~\ref{figRosa} we show the evolution of the cosmic dust yields of intermediate 
mass stars, i.e. the total dust mass produced by $<8 \, M_\odot$ stars normalized to the 
total stellar mass formed in a single instantaneous burst. The left and middle panels refer,
respectively, to $Z=10^{-4}$ and $Z=3\times 10^{-4}$. To understand the trend with metallicity
we also show in the right panel the $Z=0.001$ case \citep[][]{ventura12a, ventura12b}.
The contributions of silicate (blue dashed lines) and carbon (magenta dotted lines) dust to the total 
AGB dust budget (black dot-dashed lines) are shown. The green-thick solid lines indicate the cosmic dust 
yield\footnote{Note that dust reprocessing in the ISM (dust destruction by SN shocks and 
grain growth by accretion of heavy elements from the gas phase) is not taken into account in the present computation.} while the yellow-thin solid lines refer to the contribution from SNe, calculated taking into account the partial destruction of newly synthesized dust in the reverse shock of the
SN ($\sim 7\%$ of the newly formed dust survives; see Bianchi \& Schneider 2007 and Valiante et al. 2009 for details).

At the metallicity $Z=10^{-4}$ an important transition occurs: the cosmic dust yield is 
dominated by carbonaceous particles at all times, whereas for higher metallicities there
is an early epoch ($<200-300$ Myr) dominated by the production of silicate dust.
As shown in the figure, the epoch of transition from silicate-dominated to 
carbon-dominated dust production depends on metallicity.

In the early evolutionary stages ($100-500$ Myr) the cosmic dust yield 
in the $Z=3\times 10^{-4}$ and $Z=10^{-3}$ cases is, respectively, 3 and 7 times larger
than $Z=10^{-4}$. This trend with $Z$ is due to the behaviour of $>2 \, M_\odot$ stars,
whose dust production increases with metallicity (see Figure 5). 

At later times, the cosmic dust yields become less and less sensitive to the metallicity;
these epochs are characterized by the evolution of lower mass stars ($<2 \, M_\odot$), 
whose dust production is almost independent of $Z$ (see Figure 5). After 500 Myr the 
cosmic dust yields reach a value of $ M_{dust}/M_{star} \sim 6\times 10^{-5}$
for $Z=10^{-4}$ and $Z=3\times 10^{-4}$.

Interestingly, $\sim 1$ Gyr after the starburst, the cosmic dust yield in the $Z=0.001$ 
case is a factor of 2 smaller than for $Z=3\times 10^{-4}$ (see right-panel of 
Figure~\ref{figRosa}). This is a consequence of the lower abundance of carbon in $Z>0.001$ 
metallicity stars, as shown in the top right panel of Figure~3.

Finally, AGB stars contribution to the cosmic dust yield of the whole $[0.1-100] \, M_\odot$ 
stellar population (red dot-dashed lines), reach at most  $\sim 30\%$ (on time-scales longer 
than $\sim 2$ Gyr, close to the Hubble time at redshift $\sim 5$). Thus, metal-poor AGB stars, 
are not the main stellar dust producers, in particular at very high redshift ($z>5-6$). \\
When $Z<0.004$ (i.e. $Z<0.2 \, Z_\odot$) the dominant dust factories are higher mass stars, 
$[10-40] \, M_\odot$, exploding as SNe \citep[see][]{valiante09, valiante17}.
Conversely, metal-rich intermediate mass stars are expected to provide a significant 
\citep[or even dominant, $\sim 70\%$ on timescales of $300-500$ Myr;][]{valiante17} 
contribution to cosmic dust in galaxies (and quasars) already as redshift as high as $z>6$. As a consequence of the strongly metallicity-dependent dust production rate (and thus total dust yield) 
computed with the ATON models we expect that if galaxies are metal-poor their total dust budget is mainly 
the result of high-mass star evolution. 
On the other hand, metal-rich systems may have been significantly polluted by dust produced by AGB stars.
Our conclusion here is different from what discussed in previous 
studies \citep[see e.g.][]{valiante09, dwek11}
in which a dominant contribution from low-metallicity intermediate-mass stars is found, 
on shorter time-scales, even at $Z=0$ \citep[e.g. $>50\%$ after $300$ Myr as in][]{valiante09}.
The origin of this difference comes from the total dust yields from AGB stars adopted in these works that are only weekly dependent on stellar initial 
metallicity \citep[see e.g.][]{dwek98, zhukovska08}.

The role of AGB stars and SNe as stellar dust sources has been at the center of several studies aimed to explain 
dust enrichment of galaxies, especially at high redshifts.
To this aim, stellar dust yields are often used as an input to cosmological models which follow 
the complex process of galaxy formation and evolution. Thus, the adopted stellar yields
play a fundamental role in modelling dust (and metals) evolution through cosmic times.
The result presented here, showing a strong metallicity-dependence of the AGB star yields, is important in such a 
context. The impact of different AGB and SNe dust yields on dust production in different environments is 
discussed in \citet{gall11a}, \citet{valiante17} and \citet{ginolfi18}. 

It is important to note, however, that the (relative) role of AGB stars and SNe as stellar dust sources 
not only depends on mass- and metallicity-dependent yields \citep[][]{valiante17, ginolfi18} 
but also on the adopted IMF and SFH (more than on cosmic time, i.e. independently of 
the galaxy redshift up to $z\sim 7$)\citep{valiante09, valiante11, calura14, mancini15}. 
As an example, \citet{valiante09} or \citet{dwek11} show that AGB stars may be the dominant dust sources in dusty quasars at $z>6$ provided that a peculiar SFH is assumed. AGB stars are also shown to have a major role as stellar dust factories in metal-poor dwarf galaxies \citep[][]{zhukovska14}. 
On the other hand, different studies using AGB dust yields from \citet{fg06} or \citet{zhukovska08} conclude 
that dust enrichment in different systems (from high redshift starburst galaxies and 
quasars to local Milky Way-like and dwarf galaxies) require either a rather sizable (dominant) contribution by 
SNe, with AGB stars providing an insufficient/negligible contribution \citep[e.g.][]{michalowski10, gall11a, michalowski15}, 
an IMF biased towards high-mass stars \citep[top-heavy; e.g.][]{gall11a, gall11b, valiante11, calura14} or an additional, 
non stellar contribution 
\citep[such as grain growth or re-formation in the ISM; e.g.][]{michalowski10, valiante11, zhukovska14, michalowski15, 
zhukovska16, mancini15, ginolfi18}.
Our study suggest that these results would be further strenghtened, and in some cases pushed to their extremes 
(e.g. towards unfeasible SNe dust yields requirements), if the reduced low-metallicity AGB stars contribution 
is taken into account.

\section{Conclusions}
We present very low-metallicity models of stars of low and intermediate mass, evolving
through the AGB phase. This work extends down to metallicities $Z=10^{-4}$ previous
results from our group, limited to $Z=10^{-3}$. The models presented here have
metallicity $Z=10^{-4}$ and $Z=3\times 10^{-4}$.

We find that the mass transition to activate HBB at the base of the envelope is
$2.5~M_{\odot}$, about half solar mass lower than in stars of higher metallicity.
Low-mass stars not experiencing HBB reach the C-star stage, due to the effects of
repeated TDU events. The carbon yields and the final carbon are sensitive to the initial 
mass, while being fairly independent of metallicity. 

The strength of HBB experienced by $M \geq 2.5~M_{\odot}$ stars is higher the lower
the metallicity. The efficiency of HBB in the low-Z domain investigated here
is so strong that the surface oxygen is almost entirely destroyed by p-capture reactions
taking place at the base of the envelope; this holds particularly for the most massive
AGB stars, where the evolution of the surface chemical composition is essentially determined by HBB. 
In the $Z=10^{-4}$ case the strong destruction of oxygen makes all the stars to reach 
the C-star stage, independently of the initial mass; these stars are expected to be 
observed as C-stars during the post-AGB phases.

The dust manufactured by low-metallicity AGBs reflects the evolution of the surface
chemical composition. In the low-mass domain most of the dust is under the form of
amorphous carbon, with little ($Z=3\times 10^{-4}$) or no ($Z=10^{-4}$) traces
of silicon carbide. Despite the carbon enrichment of the surface regions is fairly
independent of metallicity, the quantity of dust produced decreases with $Z$,
because the higher effective temperatures reached by stars of lower metallicity
partly inhibits dust formation. 

In the high-mass domain the trend of dust with metallicity is more straightforward. 
These stars produce essentially silicates and alumina dust; because of the larger 
availability of silicon and aluminium in higher-metallicity stars, the dust mass 
manufactured and the size of the dust particles are higher the larger the metallicity. 
The $Z=10^{-4}$ stars follow a different behaviour, because the strong destruction of the 
surface oxygen inhibits the formation of silicates: these stars produce only carbon dust,
in negligible quantities, owing to the destruction of the surface carbon triggered 
by HBB.

In the context of cosmic dust production, the present results indicate that the AGB 
contribution to the global dust production in metal-poor environments reach
at most $\sim 30\%$ after $\sim 2$ Gyr, which corresponds to a redshift $\sim 5$. The percentages and epochs quoted above are partly dependent on the physical ingridients used to model the AGB phase, particularly to the extent of the extramixing region during the TDU (and the efficiency of the convective instability). Further sources of uncertainity are represented by the description of destruction process of the newly formed dust particles in the expanding ejecta of SNe and details of the SFH adopted. On general grounds, we
conclude that metal-poor AGB stars are not expected to be the main dust contributors, when describing low-metallicity ($Z<0.1 Z_\odot$) galaxies, at all cosmic epochs/redshifts, and that the dominant dust factorries are stars of higher mass,
exploding as supernovae. This is significantly different from the results obtained
for higher metallicities, which show that the contribution from low and intermediate
mass stars to the overall dust budget may reach $\sim 70\%$. Detailed models of dust
from AGB stars and SNe are crucial for the general understanding of stellar dust 
production in the Universe.

\section*{Acknowledgments}
The authors are indebted to the anonymous referee for the careful reading of the manuscript and for the several comments, that help improving significantly the quality of this work. FDA and DAGH acknowledge support from the State Research Agency (AEI) of the
Spanish Ministry of Science, Innovation and Universities (MCIU) and the European
Regional Development Fund (FEDER) under grant AYA2017-88254-P.
DK acknowledges support from the Macquarie University New Staff Scheme 
funding (MQNS 63822571).

\end{document}